# A new algorithm for Solving 3-CNF-SAT problem

Belal Qasemi, University of Bonab, Bonab, Iran

**belal@aut.ac.ir**

Abstract

NP-Complete problems have an important attribute that if one NP-Complete problem can be solved in polynomial time, all NP-Complete problems will have a polynomial solution. The 3-CNF-SAT problem is a NP-Complete problem and the primary method to solve it checks all values of the truth table. This task is of the $\Omega(2^n)$ time order. This paper shows that by changing the viewpoint towards the problem, it is possible to know if a 3-CNF-SAT problem is satisfiable in time $O(n^{10})$ or not?

In this paper, the value of all clauses are considered as false. With this presumption, any of the values inside the truth table can be shown in string form in order to define the set of compatible clauses for each of the strings. So, rather than processing strings, their clauses will be processed implicating that instead of $2^n$ strings, $\binom{n}{3}$ clauses are to be processed; therefore, the time and space complexity of the algorithm would be polynomial.

## 1 Introduction

The most compelling reason why theoretical computer scientists believe that P ≠ NP may be the existence of "NP-complete" problems. This class has the surprising property that if any NP-complete problem can be solved in polynomial time, then every problem in NP has a polynomial-time solution, that is, P = NP[1]. Despite years of study on this issue by different researchers, no polynomial-time algorithm has ever been discovered for any NP-complete problem.

In this paper, a novel polynomial (in time and space) algorithm is proposed to decide 3-CNF-SAT problem. The class of sat problems was shown to be NP-complete as proven in previous works [2, 3,5, 6]. The goals are to establish lower bounds in complexity, see Meyer and Sotckmeyer in [8] and Meyer in [7]. The literature in this area is rich of very nice surveys, like [4] and [9].

## 2 Problem definition

A formula such as $\varphi$ in Boolean Logic is assumed as satisfiable if there is a valuation v for set of atoms in $\varphi$ so that value($\varphi$) is true. If no such valuation exists, value($\varphi$) is false for any valuation, which means that $\varphi$ is unsatisfiable.

### 2.1 3-CNF-SAT problem

We define 3-CNF-SAT satisfiability using the following terms. A literal in a boolean formula is an occurrence of a variable or its negation. A boolean formula is in conjunctive normal form, or CNF, if it is expressed as conjunctions (by AND) of clauses, each of which is the disjunction (by OR) of one or more literals. A boolean formula is in 3-conjunctive normal form, or 3-CNF-SAT, if each clause has exactly three distinct literals. For example, the boolean formula $(x_1 \lor x_2 \lor \bar{x}_3) \land (\bar{x}_1 \lor \bar{x}_2 \lor x_3) \land (x_1 \lor \bar{x}_2 \lor x_3)$ is in 3-CNF-SAT. The first clause is $(x_1 \lor x_2 \lor \bar{x}_3)$, which contains the three literals $x_1, x_2,$ and $\bar{x}_3$.

In 3-CNF-SAT, we are asked whether a given boolean formula the φ in 3-CNF-SAT is satisfiable. The following theorem shows that a polynomial-time algorithm that can determine the satisfiability of boolean formulas is unlikely to exist, even when they are expressed in this simple normal form. satisfiability of boolean formulas in 3-conjunctive normal form is NP-complete[3].

### 2.2 The truth table

According to the Boolean Algebra for any 3-CNF formula like φ with n variables [ $x_1, x_2, ..., x_n$ ], the truth table can be formed from Table 1 in order to initialize the variables in the given formula using values in one of the rows in truth table(Table 1).

Table 1. The truth table of problem $\varphi$

| $x_1$ | $x_2$ | ... | $x_n$ | value($\varphi$) |
|---|---|---|---|---|
| *false* | *false* | | *false* | *True OR False* |
| | | ... | | ... |
| *true* | *true* | | *true* | *True OR False* |

## 2.3 Assumption

The literals of each clause are sorted based on the positional index in the list of variables.

$\forall_{clauses\ such\ as\ (l_i \vee l_j \vee l_k)}\ i < j < k$ and $l_i \in \{x_i, \bar{x}_i\}, l_j \in \{x_j, \bar{x}_j\}, l_k \in \{x_k, \bar{x}_k\}$

## 2.4 Presumption

The value of each clause is assumed as false.

$\forall_{clauses\ such\ as\ (l_i \vee l_j \vee l_k)}\ value\left((l_i \vee l_j \vee l_k)\right) = false\ so\ l_i = false,\ l_j = false,\ l_k = false, l_i \in \{x_i, \bar{x}_i\}, l_j \in \{x_j, \bar{x}_j\}, l_k \in \{x_k, \bar{x}_k\}$.

## 2.5 The table of strings

The table of strings is generated from the truth table, which holds the rows and columns of the same size. For each row of the truth table, there is only one equivalent row in the table of strings and vice versa; in fact, the table of strings is a conversion of the truth table(Table 2).

Table 2. The strings of *problem $\varphi$*

| $x_1$ | $x_2$ | ... | $x_n$ | value($\varphi$) |
|---|---|---|---|---|
| $x_1$ | $x_2$ | ... | $x_n$ | *True OR False* |
| | | ... | | ... |
| $\bar{x}_1$ | $\bar{x}_2$ | ... | $\bar{x}_n$ | *True OR False* |

---

**ALGORITHM 1**: The method to generating Table 2 based on Table 1

---

  **for** $i \leftarrow 1\ to\ 2^n$ **do**  // $n$ is the number of variables
    **for** $j \leftarrow 1\ to\ n$ **do**
      **if** $Table\ 1[i][j] = false$ **then**
        $Table\ 2[i][j] \leftarrow\ 'x_i'$ ;
      **if** $Table\ 1[i][j] = true$ **then**
        $Table\ 2[i][j] \leftarrow\ '\bar{x}_i'$ ;
    **end**
  **end**
**end**

---

*Definition 1.* Each row in the table of strings contains a string.

**LEMMA 1.** *All strings are unique.*

PROOF. The Lemma is correct due to the truth table (Table 1).

For example for $\varphi = (\bar{x}_1 \vee \bar{x}_2 \vee \bar{x}_3) \wedge (\bar{x}_2 \vee \bar{x}_3 \vee x_4) \wedge (\bar{x}_2 \vee \bar{x}_3 \vee \bar{x}_4) \wedge (x_1 \vee \bar{x}_2 \vee x_5) \wedge (\bar{x}_2 \vee x_3 \vee \bar{x}_5) \wedge (\bar{x}_1 \vee \bar{x}_2 \vee \bar{x}_6)$, truth table and string table of problem φ are in Fig. 1.

*Definition 2.* Clause_Set (w) is said to the set of all clauses resulted from the literals of one string such as w.

**ALGORITHM 2**: The method to generating Clause_Set(w)

//Suppose that w is a string like "$l_1 l_2 l_3 \ldots l_{n-1} l_n$" and the length of w is equal $n$.
$Clause\_Set(w) \leftarrow null$
**for** $i \leftarrow 1$ to $n - 2$ **do**
  **for** $j \leftarrow i + 1$ to $n - 1$ **do**
    **for** $k \leftarrow j + 1$ to $n$ **do**
      $Clause\_Set(w) \leftarrow Clause\_Set(w) \cup (l_i \vee l_j \vee l_k)$
    end
  end
end

Truth Table of φ

String Table of φ

Fig. 1 : **Truth table and string table of φ**

LEMMA 2. *lenght of Clause_Set(w) is equal $\binom{n}{3}$.*

PROOF. According to the ALGORITHM 2:

$$\text{lenght of Clause\_Set(w)} = \sum_{i=1}^{n-2}\sum_{j=i+1}^{n-1}\sum_{k=j+1}^{n} 1 = \sum_{i=1}^{n-2}\sum_{j=i+1}^{n-1}(n-j) = \sum_{i=1}^{n-2}\left((n-i-1) - \frac{1}{2}(n+i)\times(n-j-1)\right)$$

$$= n^2(n-2) - n(n-2) - n\sum_{i=1}^{n-2} i - \frac{1}{2}\left((n^2-n)(n-2) - \sum_{i=1}^{n-2} i^2 - \sum_{i=1}^{n-2} i\right)$$

$$= n^3 - 3n^2 - 2n - \frac{2n^3 - 6n^2 + 4n}{6} = \frac{n^3 - 3n^2 + 2n}{6} = \binom{n}{3}$$

**ALGORITHM 3**: The method to convert a Clause_Set(w) to string w:

$w[n] \leftarrow null$
**for** $x \leftarrow 1\ to\ \binom{n}{3}$ **do**
 **if** $(Clause\_Set(w)[x] = (l_i \vee l_j \vee l_k))$ **then**
  $w[i] \leftarrow l_i;$
  $w[j] \leftarrow l_j;$
  $w[k] \leftarrow l_k;$
 **end**
**end**

For example: $w = "x_1 x_2 x_3 \bar{x}_4 x_5 \bar{x}_6"$

Clause_Set(w) = {$(x_1 \vee x_2 \vee x_3)$, $(x_1 \vee x_2 \vee \bar{x}_4)$, $(x_1 \vee x_2 \vee x_5)$, $(x_1 \vee x_2 \vee \bar{x}_6)$, $(x_1 \vee x_3 \vee \bar{x}_4)$,
$(x_1 \vee x_3 \vee x_5)$, $(x_1 \vee x_3 \vee \bar{x}_6)$, $(x_1 \vee \bar{x}_4 \vee x_5)$, $(x_1 \vee \bar{x}_4 \vee \bar{x}_6)$, $(x_1 \vee x_5 \vee \bar{x}_6)$, $(x_2 \vee x_3 \vee \bar{x}_4)$,
$(x_2 \vee x_3 \vee x_5)$, $(x_2 \vee x_3 \vee \bar{x}_6)$, $(x_2 \vee \bar{x}_4 \vee x_5)$, $(x_2 \vee \bar{x}_4 \vee \bar{x}_6)$, $(x_2 \vee x_5 \vee \bar{x}_6)$, $(x_3 \vee \bar{x}_4 \vee x_5)$,
$(x_3 \vee \bar{x}_4 \vee \bar{x}_6)$, $(x_3 \vee x_5 \vee \bar{x}_6)$, $(\bar{x}_4 \vee x_5 \vee \bar{x}_6)$ }

LEMMA 3. *There is only one Clause_Set for each single string and each Clause_Set belongs to only one string.*

PROOF. a) The proof by reductio ad absurdum: suppose that for a string such as w there are at least two different Clause_Sets: Clause_Set₁ (w) and Clause_Set₂ (w) where Clause_Set₁ (w) ≠ Clause_Set₂ (w). As two Clause_Sets are different, so at least there is one clause which belongs to one of the two Clause_Sets and this is a contradiction because each Clause_Set contains all clauses. Therefore such a clause does not exist

b) The proof by reductio ad absurdum: suppose there is a Clause_Set which at least belongs to two strings w₁ and w₂. According to LEMMA 1, these two strings are mutually different and each string incorporates at least one different literal. Therefore Clause_Set of each of strings contains at least one clause which consists of one different literal implicating that each of the mentioned Clause_Set's consists of at least one different literal .So Clause_Set₁ (w₁) ≠ Clause_Set₂ (w₂) and this is in contradiction with the presumption of the problem.

## 2.6 The Compatibility of two clauses

Two clauses are compatible when they don't contain mutually conflicting literals but they can have different literals. For example, $(x_i \vee x_j \vee \bar{x}_k)$ and $(x_p \vee x_q \vee x_k)$ are mutually incompatible (becuse $x_k \neq \bar{x}_k$) and the clauses $(x_i \vee x_j \vee \bar{x}_k)$ and $(x_i \vee x_q \vee x_r)$ are mutually compatible ($i \neq j \neq k \neq p \neq q \neq r$).

LEMMA 4. *All clauses of one Clause_Set are compatible with each other.*

PROOF. According to the definition of Clause_Set all of its clauses don't have mutually conflicting literals so they are compatible with each other.

LEMMA 5. *Each set containing $\binom{n}{3}$ of compatible clauses resulting from n literal $l_1, l_2, \dots, l_n$ form a string.*

PROOF. All clauses are formed from literals $l_1, l_2, \dots, l_n$ and they are compatible with each other. So their forming literals can be placed beside each other in one string.

## 2.7 The clauses of peer group

According to the table of strings all strings are in the form as follows.

$$w = "l_1 l_2 \dots l_n", \forall_1^n l_i \in \{x_i, \bar{x}_i\}$$

So the prototype of all Clause_Sets is as below.

$$Clause\_Set(w) = \{(l_1 \vee l_2 \vee l_3), \ldots, (l_i \vee l_j \vee l_k)\ i < j < j, \ldots, (l_{n-2} \vee l_{n-1} \vee l_n)\}$$

Each of existing clauses in Clause_Set (w) is called the Clause prototype.

The set of clauses adherent to one given clause is called the peer group and the number of this set is 8.

For example, the prototype $(l_i \vee l_j \vee l_k)$ has the following clauses:

$$\{(x_i \vee x_j \vee x_k), (x_i \vee x_j \vee \bar{x}_k), (x_i \vee \bar{x}_j \vee x_k), (x_i \vee \bar{x}_j \vee \bar{x}_k), (\bar{x}_i \vee x_j \vee x_k), (\bar{x}_i \vee x_j \vee \bar{x}_k), (\bar{x}_i \vee \bar{x}_j \vee x_k), (\bar{x}_i \vee \bar{x}_j \vee \bar{x}_k) \mid 1 \leq i < j < k < n\}$$

LEMMA 6. *Each clause yields "true" for the values of all its peer-group clauses but for its own value it takes "false".*

PROOF. According to the presumption of the paper the value of each clause is "false" and the clauses of one group have at least one complement literal with each other. The correctness of the LEMMA 6 is proved due to the following table (Table 3).

For example : $value(c = (x_i \vee x_j \vee x_k)) = false$ then $x_i = false$ and $x_j = false$ and $x_k = false$

Table 3. An example for Lemma 6

| peer-group | valuation | value |
|---|---|---|
| $(x_i \vee x_j \vee x_k)$ | $(false \vee false \vee false)$ | $false$ |
| $(x_i \vee x_j \vee \bar{x}_k)$ | $(false \vee false \vee true)$ | $true$ |
| $(x_i \vee \bar{x}_j \vee x_k)$ | $(false \vee true \vee false)$ | $true$ |
| $(x_i \vee \bar{x}_j \vee \bar{x}_k)$ | $(false \vee true \vee true)$ | $true$ |
| $(\bar{x}_i \vee x_j \vee x_k)$ | $(true \vee false \vee false)$ | $true$ |
| $(\bar{x}_i \vee x_j \vee \bar{x}_k)$ | $(true \vee false \vee true)$ | $true$ |
| $(\bar{x}_i \vee \bar{x}_j \vee x_k)$ | $(true \vee true \vee false)$ | $true$ |
| $(\bar{x}_i \vee \bar{x}_j \vee \bar{x}_k)$ | $(true \vee true \vee true)$ | $true$ |

The algorithm processes only clauses instead of processing the values of the truth table and the table of strings. In the remainder of the paper, the CM containing all Clause_Sets are discussed.

LEMMA 7. *The number of clauses of the problem $\varphi$ with n variables is* $8 \times \binom{n}{3}$

PROOF. According to the table of strings of the problem $\varphi$, the prototype of all strings is $w = "l_1 l_2 \ldots l_n"$, $\forall_1^n l_i \in \{x_i, \bar{x}_i\}$.

Due to LEMMA 2, the number of clauses of one Clause_Set is $\binom{n}{3}$ and $Clause\_Set(w) = \{(l_1 \vee l_2 \vee l_3), \ldots, (l_i \vee l_j \vee l_k)\ i < j < j, \ldots, (l_{n-2} \vee l_{n-1} \vee l_n)\}$ So, as each of the existing clause prototypes in Clause_Set (w) can take eight different values, the number of clauses of the problem is $8 \times \binom{n}{3}$.

## 3 Data structures

### 3.1 CM (Clauses Matrix)

The Clause Matrix (CM) is a matrix incorporating all resulting clauses from all problem variables and their complements. All resulting clauses from the existing literals in problem are copied to CM. Then the existing clauses in the problem are omitted from the CM as according to the presumption of the paper, the result value of the problem for those clauses would be false. After theis removal, there may be circumstances in which some of the remaining clauses in the CM also should be deleted. If by removal of these clauses, one row of the CM become empty, the result of the problem would be false; otherwise, the presented algorithm in this paper using a data structure named as CDAG, checks the existence of the set of clauses from CM for which the result of the problem for them can be true. If the algorithm is able to create at least one CDAG from the existing clauses in the CM, it implicates that the result of the problem is true. If the algorithm could generate no CDAG, the result of the problem is false.

LEMMA 8. *The dimensions of CM is* $8 \times \binom{n}{3}$.

PROOF. It is proved according to the LEMMA 7.

**ALGORITHM 4:** The method to generation CM

GenerateCM( )
$row \leftarrow 1$  //$1 \leq row \leq \binom{n}{3}$
**for** $i \leftarrow 1$ *to* $n-2$ **do**
  **for** $j \leftarrow i+1$ *to* $n-1$ **do**
    **for** $k \leftarrow j+1$ *to* $n$ **do**
      $CM[row][1] \leftarrow (x_i \vee x_k \vee x_k)$
      $CM[row][2] \leftarrow (x_i \vee x_k \vee \bar{x}_k)$
      $CM[row][3] \leftarrow (x_i \vee \bar{x}_k \vee x_k)$
      $CM[row][4] \leftarrow (x_i \vee \bar{x}_k \vee \bar{x}_k)$
      $CM[row][5] \leftarrow (\bar{x}_i \vee x_k \vee x_k)$
      $CM[row][6] \leftarrow (\bar{x}_i \vee x_k \vee \bar{x}_k)$
      $CM[row][7] \leftarrow (\bar{x}_i \vee \bar{x}_k \vee x_k)$
      $CM[row][8] \leftarrow (\bar{x}_i \vee \bar{x}_k \vee \bar{x}_k)$
      *increment row* ;
    **end**
  **end**
**end**

$$T(n) = \sum_{i=1}^{n-2}\sum_{j=i+1}^{n-1}\sum_{k=j+1}^{n} 1 = \sum_{i=1}^{n-2}\sum_{j=i+1}^{n-1}(n-j) = \sum_{i=1}^{n-2}\left((n-i-1)-\frac{1}{2}(n+i)\times(n-j-1)\right)$$

$$= n^2(n-2) - n(n-2) - n\sum_{i=1}^{n-2}i - \frac{1}{2}\left((n^2-n)(n-2) - \sum_{i=1}^{n-2}i^2 - \sum_{i=1}^{n-2}i\right)$$

$$= n^3 - 3n^2 - 2n - \frac{2n^3 - 6n^2 + 4n}{6} = \frac{n^3 - 3n^2 + 2n}{6} = \theta(n^3)$$

LEMMA 9. *The CM covers* $Clause\_Set$ *s of all strings.*

PROOF. All clauses locate in the CM so clauses of Clause_Set of all strings are available in the CM. So, all strings can be extracted from the CM .

**ALGORITHM 5:** The method to Subtracting $\varphi$ from $CM$

**for** $i = 1$ *to* $\binom{n}{3}$ **do**
  **for** $j = 1$ *to* 8 **do**
    **If** $CM[i][j] \in \varphi$ **then**
      $CM[i][j] \leftarrow null$
  **end**
  **end**
**end**

$$T(n) = \binom{n}{3} \times 8 = \frac{n^3 - 3n^2 + 2n}{6} \times 8 = \theta(n^3)$$

LEMMA 10. *For every 3-CNF-SAT expression such as* $\varphi = c_1 \wedge \ldots \wedge c_m$, *the value of* $\varphi$ *per the clauses of* $c_m, \ldots, c_1$ *is* $false$.

PROOF. According to the Boolean algebra, the result of expressions containing the 'and' operator is false when at least one operand is $false$, so based on the convention of paper, the value of all clauses are $false$. Hence, for all clauses of $\varphi$ the result value of $\varphi$ is $false$. Based on LEMMA 10 and the paper's presumption, all clauses of the $\varphi$ are removed from the CM because the value of $\varphi$ per value of those clauses becomes $false$. Note that after subtracting $\varphi$ from $CM$ , the peer group clauses of each clause of $\varphi$ problem remain in the CM and due to LEMMA 6 the result of each clause of the $\varphi$ problem per the values of its peer group clauses becomes true.

LEMMA 11. *When a row from CM becomes empty, the result of the $\varphi$ becomes $false$.*

PROOF. 1. Based on the generation code of the CM, eight clauses like $(x_i \vee x_j \vee x_k)$ $(\bar{x}_i \vee x_j \vee \bar{x}_k)$ $(\bar{x}_i \vee x_j \vee x_k)$ $(x_i \vee x_j \vee \bar{x}_k)$ $(x_i \vee \bar{x}_j \vee x_k)$ $(x_i \vee x_j \vee \bar{x}_k)$ $(\bar{x}_i \vee \bar{x}_j \vee \bar{x}_k)$ $(\bar{x}_i \vee \bar{x}_j \vee x_k)$ exist in each row of CM.

2. Based on the generation code of the Clause_Set(w), only one of the above-mentioned clauses is available in Clause_Sets of all strings.

3. Based on LEMMA 6 per all above-mentioned clauses, the result of the $\varphi$ becomes $false$, so they are removed from the $CM$.

Based on 1, 2, and 3, the result of $\varphi$ is false per all strings

REMOVAL LEMMA 1. *By removing the clauses of $(l_i \vee l_j \vee l_k)$ and $(l_i \vee l_j \vee \bar{l}_k)$ from the CM, all clauses containing literals $l_i$ and $l_j$ become useless and should be removed. $l_i \in \{x_i, \bar{x}_i\}, l_j \in \{x_j, \bar{x}_j\}, l_k \in \{x_k, \bar{x}_k\}$.*

PROOF. According to the literals $\bar{l}_k$, $l_j$, $l_i$, $\bar{l}_j$, $\bar{l}_i$ and $l_k$, all strings are partitioned into eight following groups.

$$(l_i \vee l_j \vee l_k) \; (\bar{l}_i \vee l_j \vee \bar{l}_k) \; (\bar{l}_i \vee l_j \vee l_k) \; (l_i \vee \bar{l}_j \vee \bar{l}_k) \; (l_i \vee \bar{l}_j \vee l_k) \; (l_i \vee l_j \vee \bar{l}_k) \; (\bar{l}_i \vee \bar{l}_j \vee \bar{l}_k) \; (\bar{l}_i \vee \bar{l}_j \vee l_k)$$

All strings consist of $n$ literals with the length of $n$. By removing $(l_i \vee l_j \vee l_k)$ and $(l_i \vee l_j \vee \bar{l}_k)$, the $Clause\_Set$ of strings containing the literals of $l_j$ and $l_i$ is extracted from the CM so these strings cannot be extracted from the CM because in the k$_{th}$ place of these strings, none of the literals $l_k$ and $\bar{l}_k$ appears and their k$_{th}$ place remains empty; hence, this string will not be formed. Obviously, the $clauses$ containing the literals of $l_i$ $l_j$ only locate in the $Clause\_Set$ of these strings. Therefore these $clauses$ are useless and can be removed from the CM.

In the Table 5, the conditions of removing the clauses are tabulated.

Table 4. The patterns of removal Lemma 1

| removed clauses | Literals which make a clause useless |
|---|---|
| $(l_i \vee l_j \vee l_k)$ , $(l_i \vee l_j \vee \bar{l}_k)$ | $l_i \, l_j$ |
| $(l_i \vee l_j \vee l_k)$ , $(l_i \vee \bar{l}_j \vee l_k)$ | $l_i \, l_k$ |
| $(l_i \vee l_j \vee l_k)$ , $(\bar{l}_i \vee l_j \vee l_k)$ | $l_j \, l_k$ |
| $(l_i \vee \bar{l}_j \vee \bar{l}_k)$ , $(\bar{l}_i \vee \bar{l}_j \vee \bar{l}_k)$ | $\bar{l}_j \, \bar{l}_k$ |
| $(\bar{l}_i \vee l_j \vee \bar{l}_k)$ , $(\bar{l}_i \vee \bar{l}_j \vee \bar{l}_k)$ | $\bar{l}_i \, \bar{l}_k$ |
| $(\bar{l}_i \vee \bar{l}_j \vee l_k)$ , $(\bar{l}_i \vee \bar{l}_j \vee \bar{l}_k)$ | $\bar{l}_i \, \bar{l}_j$ |

REMOVAL LEMMA 2. *By removing the clauses of $(l_i \vee l_j \vee l_k)$, $(l_i \vee l_j \vee \bar{l}_k)$, $(l_i \vee \bar{l}_j \vee \bar{l}_k)$ and $(l_i \vee \bar{l}_j \vee l_k)$ from the CM, all clauses containing literal $l_j$ become useless and should be removed.*

PROOF. According to the literals $\bar{l}_k$, $l_j$, $l_i$, $\bar{l}_j$, $\bar{l}_i$ and $l_k$, all strings are partitioned in eight following groups.

$$(l_i \vee l_j \vee l_k) \; (\bar{l}_i \vee l_j \vee \bar{l}_k) \; (\bar{l}_i \vee l_j \vee l_k) \; (l_i \vee \bar{l}_j \vee \bar{l}_k) \; (l_i \vee \bar{l}_j \vee l_k) \; (l_i \vee l_j \vee \bar{l}_k) \; (\bar{l}_i \vee \bar{l}_j \vee \bar{l}_k) \; (\bar{l}_i \vee \bar{l}_j \vee l_k)$$

All strings consist of $n$ literals with the length of $n$. By removing $(l_i \vee l_j \vee l_k)$, $(l_i \vee l_j \vee \bar{l}_k)$, $(l_i \vee \bar{l}_j \vee \bar{l}_k)$ and $(l_i \vee \bar{l}_j \vee l_k)$, the $Clause\_Set$ of strings containing the literal $l_i$ does not come out from the CM so these strings cannot be extracted from the CM because in the k$_{th}$ and j$_{th}$ places of these strings none of the literals $l_j l_k$, $l_j \bar{l}_k$, $\bar{l}_j \bar{l}_k$ and $\bar{l}_j l_k$ appears and their k$_{th}$ and j$_{th}$ places remains empty. Hence, these strings will not be formed. Obviously, the *clauses* containing the literals of $l_i$ only locate in the *Clause_Set* of these strings. Therefore, these *clauses* are useless and can be removed from the CM.

In the Table 6, the removal conditions of REMOVAL LEMMA 2 are tabulated.

Table 5. The patterns of removal Lemma 2

| removed clauses | Literals which make a clause useless |
|---|---|
| $(l_i \vee l_j \vee l_k), (l_i \vee l_j \vee \overline{l_k}), (l_i \vee \overline{l_j} \vee l_k), (l_i \vee \overline{l_j} \vee \overline{l_k})$ | $l_i$ |
| $(\overline{l_i} \vee l_j \vee l_k), (\overline{l_i} \vee l_j \vee \overline{l_k}), (\overline{l_i} \vee \overline{l_j} \vee l_k), (\overline{l_i} \vee \overline{l_j} \vee \overline{l_k})$ | $\overline{l_i}$ |
| $(l_i \vee l_j \vee l_k), (l_i \vee l_j \vee \overline{l_k}), (\overline{l_i} \vee l_j \vee l_k), (\overline{l_i} \vee l_j \vee \overline{l_k})$ | $l_j$ |
| $(l_i \vee \overline{l_j} \vee l_k), (\overline{l_i} \vee \overline{l_j} \vee l_k), (l_i \vee \overline{l_j} \vee \overline{l_k}), (\overline{l_i} \vee \overline{l_j} \vee \overline{l_k})$ | $\overline{l_j}$ |
| $(l_i \vee l_j \vee l_k), (\overline{l_i} \vee l_j \vee l_k), (l_i \vee \overline{l_j} \vee l_k), (\overline{l_i} \vee \overline{l_j} \vee l_k)$ | $l_k$ |
| $(l_i \vee l_j \vee \overline{l_k}), (\overline{l_i} \vee l_j \vee \overline{l_k}), (l_i \vee \overline{l_j} \vee \overline{l_k}), (\overline{l_i} \vee \overline{l_j} \vee \overline{l_k})$ | $\overline{l_k}$ |

## 3.2 PRC (Public Removal Conditions)

The PRC is a matrix with the dimensions of $2n * 2n$ trueisfying the removal conditions related to the $CM$.

$if (PRC[i,i] = 1)$

any clause has $l_i$ is useless, $l_i \in \{x_i, \bar{x}_i\}$

$if (PRC[i,j] = 1)$

any clause has $l_i$ and $l_j$ is useless, $l_i \in \{x_i, \bar{x}_i\}, l_j \in \{x_j, \bar{x}_j\}$

LEMMA 12. *If after removal operation, one row from CM remains empty, then the result of the problem $\varphi$ becomes false.*

PROOF. The removal operation deletes those clauses that are the member of the *Clause_Set*, the equivalent string of which, falsifies the problem $\varphi$. Based on the generation code of the CM, eight clauses exist in each row of this matrix:

$(x_i \vee x_j \vee x_k) (\bar{x}_i \vee x_j \vee \bar{x}_k) (\bar{x}_i \vee x_j \vee x_k) (x_i \vee \bar{x}_j \vee \bar{x}_k) (x_i \vee \bar{x}_j \vee x_k) (x_i \vee x_j \vee \bar{x}_k) (\bar{x}_i \vee \bar{x}_j \vee \bar{x}_k) (\bar{x}_i \vee \bar{x}_j \vee x_k)$. In Clause_Set of all strings only one of the aforementioned clauses is available. Therefore the removal of all these clauses culminates at the $\varphi$ being $false$ per all strings. Hence the result of $\varphi$ is false.

LEMMA 13. *The result of problem $\varphi$ is true when there exists at least one $Clause\_Set$ in the CM.*

PROOF. If one *Clause_Set* is resulted from the CM, based on LEMMA 6, the result of all *clauses* of the problem φ per the values of existing clauses in the outcome Clause_Set are true because these clauses are peer-group with clauses of the problem φ so the result of the problem φ is true.

**ALGORITHM 6:** The method to finding removal conditions

**FindRemovalConditions** (**Matrix**[][] $matrix$, **RemovalConditions**[][] $rc$)
$newRemovalCondition \leftarrow false$
**for** $i \leftarrow 1\ to\ \binom{n}{3}$ **do**
//Assume $matrix[i][1] = (x_p \vee x_q \vee x_r), matrix[i][2] = (x_p \vee x_q \vee \bar{x}_r),$
//$matrix[i][3] = (x_p \vee \bar{x}_q \vee x_r), matrix[i][4] = (x_p \vee \bar{x}_q \vee \bar{x}_r),$
//$matrix[i][5] = (\bar{x}_p \vee x_q \vee x_r), matrix[i][6] = (\bar{x}_p \vee x_q \vee \bar{x}_r),$
//$matrix[i][7] = (\bar{x}_p \vee \bar{x}_q \vee x_r), matrix[i][8] = (\bar{x}_p \vee \bar{x}_q \vee \bar{x}_r),$
//$1 \le p < q < r \le n$

**if**$(matrix[i][1] = null$ **and** $matrix[i][2] = null$ **and** $rc[\textbf{index }(x_p)][\textbf{ index }(x_q)] = 0)$ **then**
    $rc[\textbf{index }(x_p)][\textbf{ index }(x_q)] \leftarrow 1$
    $newRemovalCondition \leftarrow true$
**end**
**if**$(matrix[i][1] = null$ **and** $matrix[i][3] = null$ **and** $rc[\textbf{index }(x_p)][\textbf{ index }(x_r)] = 0)$ **then**
    $rc[\textbf{index }(x_p)][\textbf{ index }(x_r)] \leftarrow 1$
    $newRemovalCondition \leftarrow true$
**end**
**if**$(matrix[i][1] = null$ **and** $matrix[i][5] = null$ **and** $rc[\textbf{index }(x_q)][\textbf{ index }(x_r)] = 0)$ **then**
    $rc[\textbf{index }(x_q)][\textbf{ index }(x_r)] \leftarrow 1$
    $newRemovalCondition \leftarrow true$
**end**
**if**$(matrix[i][2] = null$ **and** $matrix[i][4] = null$ **and** $rc[\textbf{index }(x_p)][\textbf{ index }(\bar{x}_r)] = 0)$ **then**
    $rc[\textbf{index }(x_p)][\textbf{ index }(\bar{x}_r)] \leftarrow 1$
    $newRemovalCondition \leftarrow true$
**end**
**if**$(matrix[i][2] = null$ **and** $matrix[i][6] = null$ **and** $rc[\textbf{index }(x_q)][\textbf{ index }(\bar{x}_r)] = 0)$ *then*
    $rc[\textbf{index }(x_q)][\textbf{ index }(\bar{x}_r)] \leftarrow 1$
    $newRemovalCondition \leftarrow true$
**end**
**if**$(matrix[i][3] = null$ **and** $matrix[i][4] = null$ **and** $rc[\textbf{index }(x_p)][\textbf{ index }(\bar{x}_q)] = 0)$ **then**
    $rc[\textbf{index }(x_p)][\textbf{ index }(\bar{x}_q)] \leftarrow 1$
    $newRemovalCondition \leftarrow true$
**end**
**if**$(matrix[i][3] = null$ **and** $matrix[i][7] = null$ **and** $rc[\textbf{index }(\bar{x}_q)][\textbf{ index }(x_r)] = 0)$ **then**
    $rc[\textbf{index }(\bar{x}_q)][\textbf{ index }(x_r)] \leftarrow 1$
    $newRemovalCondition \leftarrow true$
**end**
**if**$(matrix[i][4] = null$ **and** $matrix[i][8] = null$ **and** $rc[\textbf{index }(\bar{x}_q)][\textbf{ index }(\bar{x}_r)] = 0)$ **then**
    $rc[\textbf{index }(\bar{x}_q)][\textbf{ index }(\bar{x}_r)] \leftarrow 1$
    $newRemovalCondition \leftarrow true$
**end**
**if**$(matrix[i][5] = null$ **and** $matrix[i][7] = null$ **and** $rc[\textbf{index }(\bar{x}_p)][\textbf{ index }(x_r)] = 0)$ **then**
    $rc[\textbf{index }(\bar{x}_p)][\textbf{ index }(x_r)] \leftarrow 1$
    $newRemovalCondition \leftarrow true$
**end**
**if**$(matrix[i][6] = null$ **and** $matrix[i][8] = null$ **and** $rc[\textbf{index }(\bar{x}_p)][\textbf{ index }(\bar{x}_r)] = 0)$ **then**
    $rc[\textbf{index }(\bar{x}_p)][\textbf{ index }(\bar{x}_r)] \leftarrow 1$
    $newRemovalCondition \leftarrow true$
**end**
**if**$(matrix[i][7] = null$ **and** $matrix[i][8] = null$ **and** $rc[\textbf{index }(\bar{x}_p)][\textbf{ index }(\bar{x}_q)] = 0)$ **then**
    $rc[\textbf{index }(\bar{x}_p)][\textbf{ index }(\bar{x}_q)] \leftarrow 1$
    $newRemovalCondition \leftarrow true$
**end**
**if**$(matrix[i][1] = null$ **and** $matrix[i][2] = null$ **and** $matrix[i][3] = null$ **and** $matrix[i][4]$
        $= null$ **and** $rc[\textbf{index }(x_p)][\textbf{ index }(x_p)] = 0)$ **then**
    $rc[\textbf{ index }(x_p)][\textbf{ index }(x_p)] \leftarrow 1$
    $newRemovalCondition \leftarrow true$
**end**
**if**$(matrix[i][1] = null$ **and** $matrix[i][2] = null$ **and** $matrix[i][5] = null$ **and** $matrix[i][6]$
        $= null$ **and** $rc[\textbf{index }(x_q)][\textbf{ index }(x_q)] = 0)$ **then**
    $rc[\textbf{ index }(x_q)][\textbf{ index }(x_q)] \leftarrow 1$
    $newRemovalCondition \leftarrow true$
**end**
**if**$(matrix[i][1] = null$ **and** $matrix[i][3] = null$ **and** $matrix[i][5] = null$ **and** $matrix[i][7]$
        $= null$ **and** $rc[\textbf{index }(x_r)][\textbf{ index }(x_r)] = 0)$ **then**
    $rc[\textbf{ index }(x_r)][\textbf{ index }(x_r)] \leftarrow 1$
    $newRemovalCondition \leftarrow true$
**end**
**if**$(matrix[i][5] = null$ **and** $matrix[i][6] = null$ **and** $matrix[i][7] = null$ **and** $matrix[i][8]$
        $= null$**and** $rc[\textbf{ index }(\bar{x}_p)][\textbf{ index }(\bar{x}_p)] = 0)$ **then**

$rc[\text{ index }(\bar{x}_p)][\text{ index }(\bar{x}_p)] \leftarrow 1$
$newRemovalCondition \leftarrow true$
**end**
**if**$(matrix[i][3] = null$ **and** $matrix[i][4] = null$ **and** $matrix[i][7] = null$ **and** $matrix[i][8]$
$= null$ **and** $rc[\text{index }(\bar{x}_q)][\text{ index }(\bar{x}_q)] = 0)$ **then**
$rc[\text{ index }(\bar{x}_q)][\text{ index }(\bar{x}_q)] \leftarrow 1$
$newRemovalCondition \leftarrow true$
**end**
**if**$(matrix[i][2] = null$ **and** $matrix[i][4] = null$ **and** $matrix[i][6] = null$ **and** $matrix[i][8]$
$= null$ **and** $rc[\text{index }(x_p)][\text{ index }(x_q)] = 0)$ **then**
$rc[\text{ index }(\bar{x}_r)][\text{ index }(\bar{x}_r)] \leftarrow 1$
$newRemovalCondition \leftarrow true$
**end**
**end** // end of for
**Return** $newRemovalCondition$
**end** //end of algorithm

$$T(n) = \sum_{1}^{\binom{n}{3}} 17 = 17 \times \binom{n}{3} = 17 \times \frac{n^3 - 3n^2 + 2n}{6} = \theta(n^3)$$

**ALGORITHM 7:** The method to finding index of a literal in Removal condition matrix

**index (literal $l_i$)**
  **if** ($l_i = x_i$) **then**
    **return** $2 \times i - 1$
  **else if** ($l_i = \bar{x}_i$) **then**
    **return** $2 \times i$
**end**

**ALGORITHM 8:** The method to garbage collection

**Garbage collection(Matrix[][]** $matrix$, **RemovalConditions[][]** $rc$)
**foreach** $Clause\ c$ **in** $matrix$
     **if**($c\ includes\ Removal\ condition$) **then**
         $Remove\ c\ from\ matrix$ // $matrix[i][j] \leftarrow null$
$newRemovalCondition \leftarrow$ **FindRemovalConditions**$(matrix, rc)$
**if** ($newRemovalCondition = true$) **then**
**Garbage collection**$(matrix, rc)$
**else if**($newRemovalCondition = false$) **then**
     $exite$
**end** // end of algorithm

$$T(n) = \text{ all removal conditions} \times \text{all clauses in matrix} = (2 \times n)^2 \times 8 \times \binom{n}{3} = 4n^2 \times 8 \times \frac{n^3 - 3n^2 + 2n}{6} = O(n^5)$$

**ALGORITHM 9**: The method to checking matrix

**MatrixIsValid(Matrix[][]** $matrix$)
**foreach** $Row\ like\ row$ **in** $matrix$
   **if**($matrix[row][]\ is\ empty$) **then**
       **Return** $false$
**end**
**Return** $true$
**end**

$$T(n) = \text{The number of rows of the matrix} = \binom{n}{3} = \frac{n^3 - 3n^2 + 2n}{6} = O(n^3)$$

## 3.3 CDAG (Clauses Directed Acyclic Graph)

After subtracting φ from the CM, those clauses that make φ false as well as useless Clauses are removed from the CM. Based on LEMMA 13, if at least one Clause_Set results from CM, the result value of the problem φ is true. The suggested algorithm uses the CDAG data structure in order to check the existence or absence of *Clause_Set* in the CM. The root of the CDAG is equal to one of the *clauses* of the first row in the CM. Due to the number of the *clauses* in the first row of the CM, at least eight the CDAG's can be created for the problem φ .In the following part, we will prove that if at least one the CDAG is successfully generated, the result of the problem φ is true.

### 3.3.1 The structure of CDAG

The CDAG is a directed acyclic graph which is implemented by a doubly linked list. The CDAG's nodes contain one *clause* and two links of *left* and *right* for the left and right children. The CDAG consists of $\binom{n}{3}$ columns with at least one node and at most eight nodes. In the first column of the CDAG, a node exists with the name of "root" which contains one *clause* from the first row of the CM. In the last column of the CDAG, two leaves exist at most. The nodes of the CDAG's $k_{th}$ column ($1 \leq k < \binom{n}{3}$) have at least one and at most two chilren in k+1th column except for the leaves which have no children.

When there is an edge between two nodes containing $c_1$ and $c_2$, it means that the *clauses* of $c_1$ and $c_2$ are compatible.

Each the CDAG incorporates n-3 sections as $l_n, l_{n-1}, \ldots, l_5, l_4$ (see Fig. 2).

The section $l_i$ consists of $(n-2-i)$ subsections $l_i l_n, l_i l_{n-1}, \ldots, l_i l_{i+1}$ .There exist at least one and at most four edges between two sections $l_p$ and $l_q$,; and due to the REMOVAL LEMMA 1 in the columns where the *clauses* related to $l_p$ and $l_q$ exist, the combination of $l_p l_q$ would also exist. The literal $l_i$ exits in the *clauses* of sections 4... , but it does not exist in sections $i+1, \ldots, n$ .

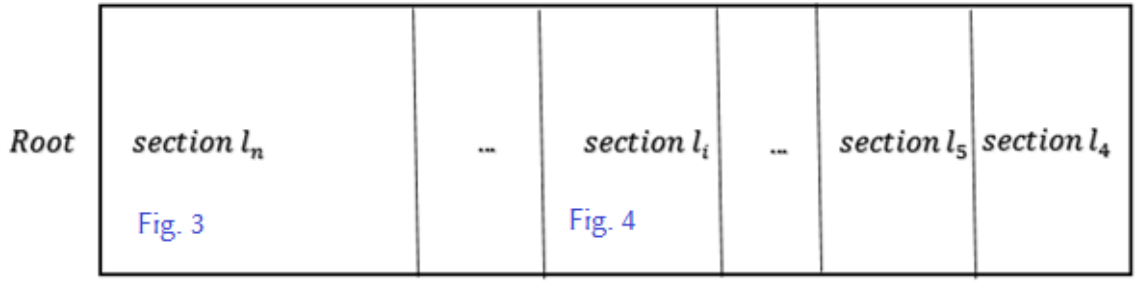

**Fig. 2. Sections of CDAG**

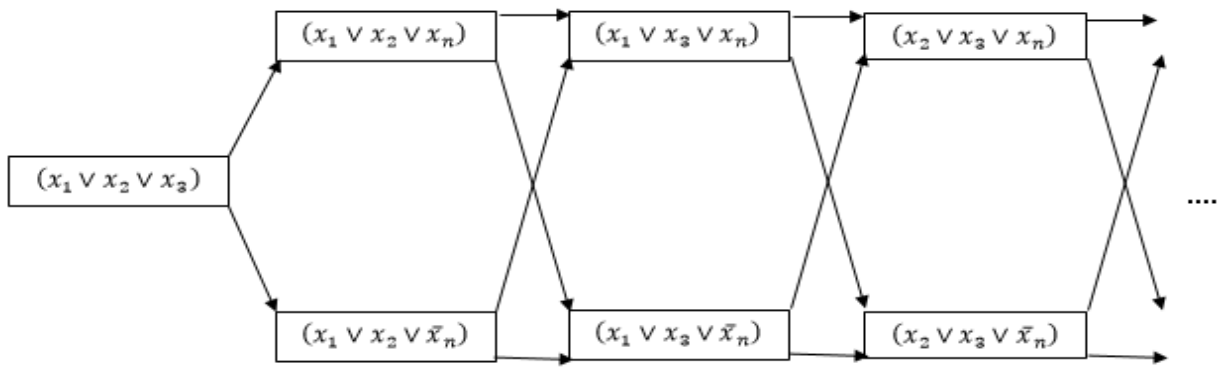

**Fig. 3. Section $l_n$**

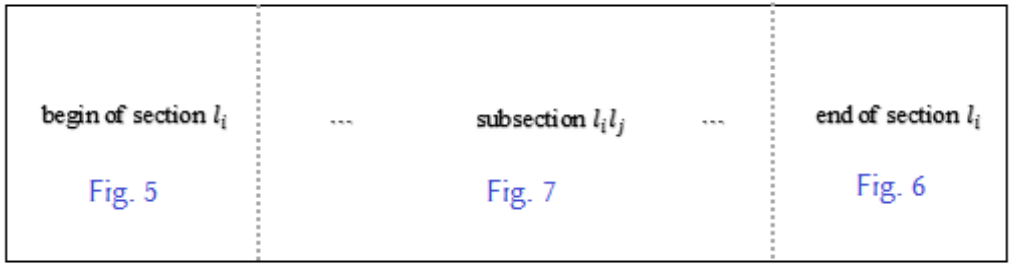

**Fig. 4. Section $l_i$**

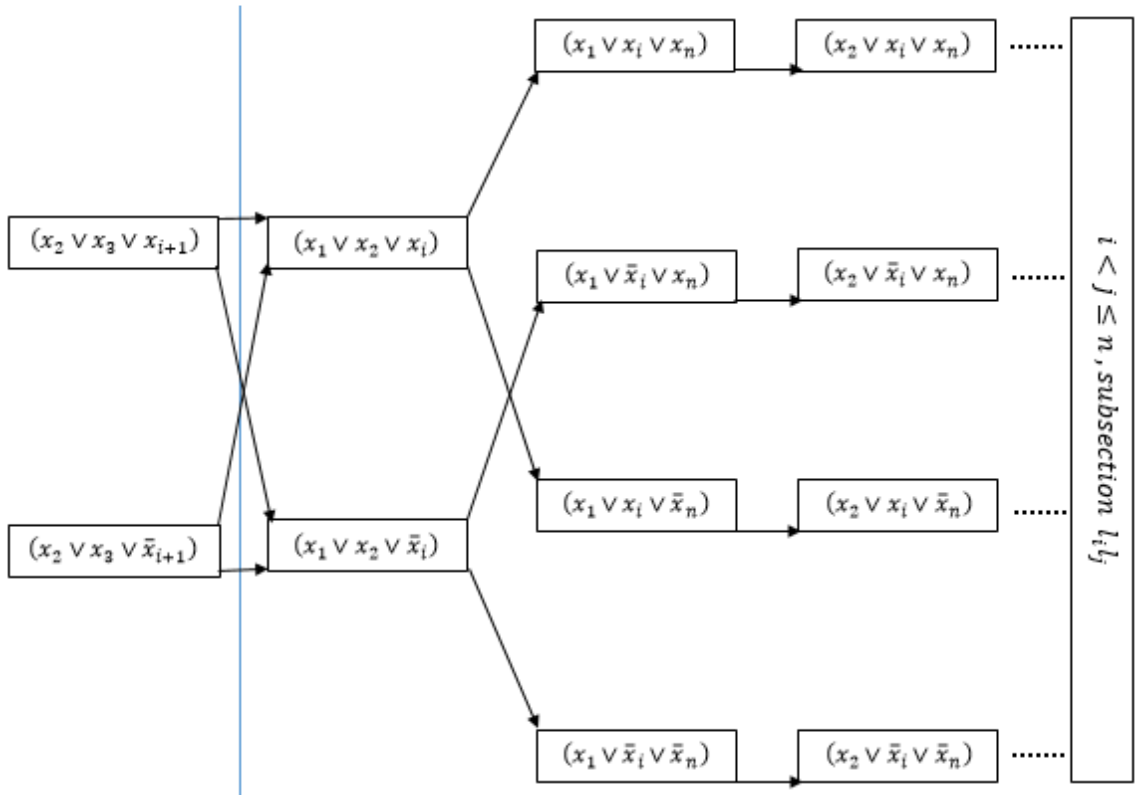

**Fig. 5. Begining of section $l_i$**

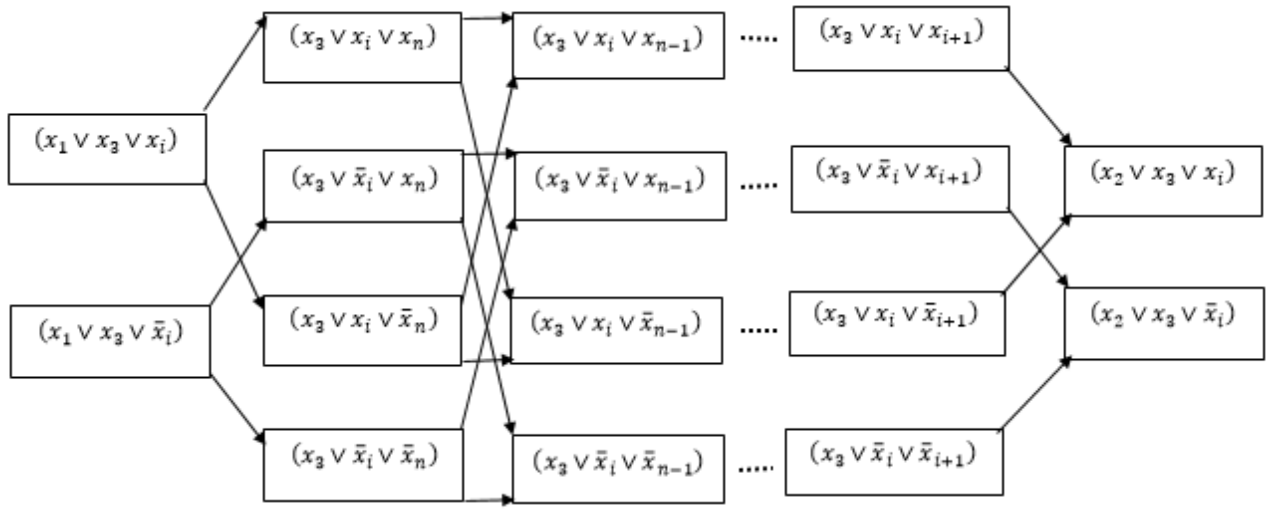

**Fig. 6. End of section $l_i$**

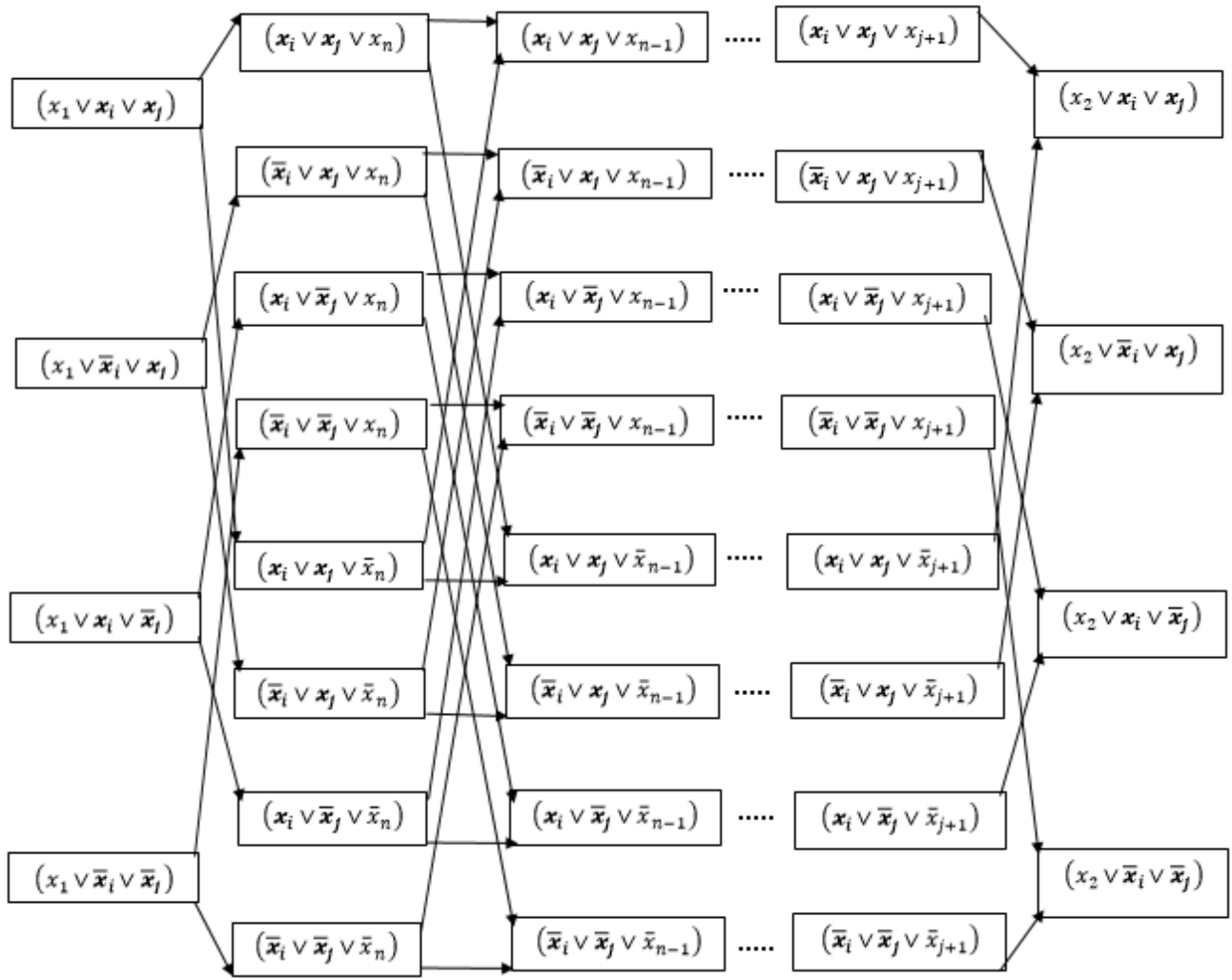

**Fig. 7. Subsection $l_i l_j$**

Section $l_i$ consists of all *clauses* containing $x_i$ and $\bar{x}_i$, so this section is divided into two separate sets of $x_i$ and $\bar{x}_i$. In $x_i$, the set of *clauses* containing $x_i$ is locate and in the $\bar{x}_i$, set the *clauses* containing $\bar{x}_i$. In section $l_i$, at least one of the $x_i$ and $\bar{x}_i$ sections exist. Each column of the CDAG relates to the clauses of one group such as $(l_i \vee l_j \vee l_k)$ where: $(1 \le i < j < k \le n)$ and $l_i \in \{x_i, \bar{x}_i\}, l_j \in \{x_j, \bar{x}_j\}, l_k \in \{x_k, \bar{x}_k\}$.

LEMMA 14. *If in the CDAG, the variables $x_p$ and $\bar{x}_p$ exist, in the columns containing the clauses related to the literal $l_p$ ($l_p \in \{x_p, \bar{x}_p\}$), the clauses containing the variables of $x_p$ and $\bar{x}_p$ would definitely exist.*

PROOF. It is true due to the REMOVAL LEMMA 2.

LEMMA 15. *If in the CDAG, the combination of $x_p x_q$ exists then in the columns related to the literals containing the combination of $l_p l_q$ ($l_p \in \{x_p, \bar{x}_p\}, l_q \in \{x_q, \bar{x}_q\}$) the clauses containing the combination of $x_p x_q$ would definitely exist.*

PROOF. It is true due to the REMOVAL LEMMA 1.

### 3.3.2 SM (Source Matrix), LRC (Local Removal Conditions)

For generating the CDAG, two data structures of LRC and SM are required.

### 3.3.2.1 SM

The SM is a matrix with the same size of the CM containing all of its clauses, which can be used in the construction of the CDAG. The initialization of SM works as follows:

---

**ALGORITHM 10:** The method to generating SM

**GeneratingSM**(integer *column*)
   $SM[1][column] \leftarrow CM[1][column]$
   **for** $i = 2$ *to* $\binom{n}{3}$ **do**
      **for** $j = 1$ *to* $8$ **do**
         **if**($CM[i][j] \ne null$ **and** $CM[1][column]$
            **and** $CM[i][j]$ *are compatible* ) **then**
               $SM[i][j] \leftarrow CM[i][j]$
   **end** // end of for(i)
   **if**(**MatrixIsValid**($SM$) = $true$) **then**
      $newRemovalCondition \leftarrow$ **FindRemovalConditions**($SM, LRC$)
      **if**($newRemovalCondition = true$) **then**
         Garbage collection($SM, LRC$)
      **else if**($newRemovalCondition = false$) **then**
         exite
   **end** // end of if (MatrixIsValid)
   **else if**(**MatrixIsValid**($SM$) = $false$) **then**
      exite
**end** // end of algorithm

---

$$T(n) = 8 \times \binom{n}{3} + O(n^5) = 8 \times \frac{n^3 - 3n^2 + 2n}{6} + O(n^5) = O(n^5)$$

### 3.3.2.2 LRC

The LRC is a data structure in PRC dimensions containing all the CDAG removal conditions. The LRC is initialized by the PRC values.
The initialization of LRC works as follows:

---

**ALGORITHM 11:** The method to generating LRC

**GeneratingLRC**()
   $LRC[2n][2n] \leftarrow PRC$
   **FindRemovalConditions** ($SM, LRC$)
**end**

---

$T(n) = (2n)^2 = O(n^2)$

## The construction of CDAG:

In the first stage, the root of the CDAG is selected among existing clauses of the first row. Assume that $c = (x_1 \vee x_2 \vee x_3)$ is selected. By choosing the root, new conditions of removal emerge and these conditions are entered into the LRC. Then, all compatible clauses with the root inside the CM, which lack removal requirements are entered into the SM. In case of the birth of any new removal condition, that new condition is entered into the LRC. If a row from the SM is not empty, the construction of CDAG goes on as below:

The construction of the CDAG is done in stage $n - 4$. The CDAG generated in stage i ($4 \leq i \leq n$), is named $CDAG_i$. In stage i, two clauses of $c = (x_1 \vee x_2 \vee x_i)$ and $\bar{c} = (x_1 \vee x_2 \vee \bar{x}_i)$ are inserted into the CDAG given that they are available in the SM.

$$CDAG_1 \leftarrow CDAG_i, CDAG_2 \leftarrow CDAG_i$$

"c" is inserted into the $CDAG_1$ and "$\bar{c}$" is inserted to $CDAG_2$.

**The method of inserting c to the $CDAG_1$:**

The $CDAG_1$ is scanned from the first column to the last one. In column $i$ ($CDAG_1.Length \geq i \geq 1$), the combination of c with the clauses of the nodes of that column is orderly computed and for each combination, it is checked if that combination is compatible with the $CDAG_1$ or not. and also if it has no removal condition. Then, all resulted clauses from that combination are attached to the combined node. If c is not added to any one of the nodes in the i$_{th}$ column, c is not added to the $CDAG_1$.

**ALGORITHM 12:** The method to insert clause c to $CDAG_1$

**InsertToCDAGc**(**clause** $c$, **CDAG** $CDAG_1$)
$count \leftarrow CDAG_1.length$
$A \leftarrow CDAG_1.root$
**while**($A \neq null$ **and** $count > 0$) **do**
  $B \leftarrow null$
  **for** $i \leftarrow 1$ to $A.length$ **do**
    $D \leftarrow \{all\ littrals\ of\ A[i].clause\} \cup \{all\ littrals\ of\ c\}$
    **Sort**($D$)
    **if**(**CompatibleDwithCDAG**($D, CDAG_1, A[i]$) = $true$) **then**
      $List \leftarrow$ **ListClausesOfD**($D$)
      $B \leftarrow B \cup A[i].left$
      $B \leftarrow B \cup A[i].right$
      $t \leftarrow A[i].left$
      $s \leftarrow A[i].right$
      $r \leftarrow A[i]$
      **for** $j \leftarrow 1$ to $List.lenght$ **do**
        **if**($List[j] \notin SM$) **then**
          **Return** false
        **if**($List[j] \notin CDAG_1$) **then**
          $r.left \leftarrow List[j]$
          $r \leftarrow r.left$
        **end**
      **end** // end of for(j)
      $r.left \leftarrow t$
      $r.right \leftarrow s$
    **end**
  **end**
  $A \leftarrow B$
  *Decriment Count*
**end** // end of for
**end** // end of while($A \neq null$ and $count > 0$)
**if**($count > 0$) **then**
  **Return** $false$
**else if** ($count = 0$) **then**
  **Return** $true$

**end** // end of InsertToCDAG**c**

$T(n)$ = length of CDAG $\times$ $T(n)$ of CompatibleDwithCDAG = $O(n^3) \times O(n^3) = O(n^6)$

**The method of inserting $\bar{c}$ to the $CDAG_2$:**

The $CDAG_2$ is scanned from the first column to the last one. In column $i$ ($CDAG_2.Length \geq i \geq 1$), the combination of $\bar{c}$ with the clauses of the nodes of that column is orderly computed and for each combination, it is checked if that combination is compatible with the $CDAG_2$ or not. and also if it has no removal condition. Then, all resulted clauses from that combination are attached to the combined node. If $\bar{c}$ is not added to any one of the nodes in the i$_{th}$ column, $\bar{c}$ is not added to the $CDAG_2$.

**ALGORITHM 13:** The method to insert clause $\bar{c}$ to $CDAG_2$

**InsertToCDAG$\bar{c}$**(**clause** $\bar{c}$, **CDAG** $CDAG_2$)
$count \leftarrow CDAG_2.length$

```
A ← CDAG₂.root
while(A ≠ null and count > 0) do
   B ← null
   for i ← 1 to A.length do
      D ← {all littrals of A[i].clause} ∪ {all littrals of c}
      Sort(D)
      if(CompatibleDwithCDAG(D, CDAG₂, A[i]) = true) then
         List ← ListClausesOfD(D)
         B ← B ∪ A[i].left
         B ← B ∪ A[i].right
         t ← A[i].left
         s ← A[i].right
         r ← A[i]
         for j ← 1 to List.lenght do
            if(List[j] ∉ SM) then
               Return false
            if(List[j] ∉ CDAG₂) then
               r.right ← List[j]
               r ← r.right
            end
         end // end of for(j)
         r.left ← t
         r.right ← s
      end
   end
   A ← B
   Decriment Count
end // end of for
end // end of while(A ≠ null and count > 0)
if(count > 0) then
   Return false
else if (count = 0) then
   Return true
end // end of InsertToCDAGc̄
```

T(n) = length of CDAG × T(n) of CompatibleDwithCDAG = O(n³) × O(n³) = O(n⁶)

**ALGORITHM 14:** The method of compatibility check for D and the CDAG

```
CompatibleDwithCDAG(littral D[], CDAG cdag, Node currentNode)
P ← Cdag.root
While(D does not have any removal condition and currentNode ∉ P and P ≠ null) do
Q ← null
for i ← 1 to P.length do
   if(D and P[i].clauses are compatible) then
      Q ← Q ∪ P[i].left
      Q ← Q ∪ P[i].right
   end
end
P ← null
P ← Q
end // end of while
if(currentNode ∈ P) then
   Return true
else
   Return false
end // end of CompatibleDwithCDAG
```

$T(n) = \text{length of CDAG} = \binom{n}{3} = \frac{n^3 - 3n^2 + 2n}{6} = O(n^3)$

**ALGORITHM 15:** The method to generating list of clauses of D

**ListClausesOfD(literal $D[]$)**
 $List \leftarrow null$
 **for** $i \leftarrow 1\ to\ D.length - 2$ **do**
   **for** $j \leftarrow i + 1\ to\ D.length - 1$ **do**
     **for** $k \leftarrow j + 1\ to\ D.length$ **do**
       $list \leftarrow list \cup (l_i \vee l_j \vee l_k)\ //l_i, l_j, l_k \in D$
 **Return** $list$
**end**

$$T(n) = \sum_{i=1}^{D.length-2}\sum_{j=i+1}^{D.length-1}\sum_{k=j+1}^{D.length} 1 = \frac{D.length^3 - 3D.length^2 + 2D.length}{6} \xrightarrow{D.length \le 6} T(n) \le \binom{6}{3} = O(1)$$

After inserting "c" into the $CDAG_1$ and "$\bar{c}$" into the $CDAG_2$, $CDAG_i$ is generated as follows:

$CDAG_i \leftarrow Merg(CDAG_1, CDAG_2)$

**ALGORITHM 16:** The method to merging $CDAG_1$ and $CDAG_2$

**Merge(CDAG $CDAG_1$, CDAG $CDAG_2$)**
//if n ∈ $CDAG_1$ and m ∈ $CDAG_2$ and n.cluse = m.cluse, then m = n.
**if**($CDAG_1 \ne null$ **and** $CDAG_2 = null$) **then**
   **Return** $CDAG_1$
**else if**($CDAG_1 = null$ **and** $CDAG_2 \ne null$) **then**
       **Return** $CDAG_2$
**else if**($CDAG_1 \ne null$ **and** $CDAG_2 \ne null$) **then**
       $A \leftarrow CDAG_1.root$
       $B \leftarrow CDAG_2.root$
       $Cdag.root \leftarrow CDAG_1.root$
       $P \leftarrow A \cup B$
     **while**($P \ne null$) **do**
           $Q \leftarrow null$
           $C \leftarrow null$
           $D \leftarrow null$
         **for** $i \leftarrow 1\ to\ A.length$ **do**
             $C \leftarrow C \cup A[i].left \cup A[i].right$ // A navigates the $CDAG_1$
         **for** $j \leftarrow 1\ to\ B.length$ **do**
             $D \leftarrow D \cup B[j].left \cup B[j].right$ //B navigates the $CDAG_2$
         $Q \leftarrow C \cup D$
         **if**($Q \ne null$) **then**
            **for** $i \leftarrow 1\ to\ Q.length$ **do**
                $Cdag.add(Q[i])$
            **for** $j \leftarrow 1\ to\ P.length$ **do**
                $P[j].left \leftarrow A[j].left \cup B[j].left$
                $P[j].right \leftarrow A[j].right \cup B[j].right$
            **end** // end of for(j)
         **end** // end of if
         $A \leftarrow C$
         $B \leftarrow D$
         $P \leftarrow Q$
      **end**//end of while
   **end** // end of else if
   **Return** $Cdag$
**end**

$T(n) = $ length of CDAG $= \binom{n}{3} = \frac{n^3 - 3n^2 + 2n}{6} = O(n^3)$

If none of the clauses "c" and "c̄" is added to the $CDAG_i$, the construction of the $CDAG_i$ fails and the existing clause in the root of the CDAG is removed from the CM and in case of observing any new removal condition, this condition is added into the PRC and then CM is updated.

After the creation of the $CDAG_i$, Any garbage will be removed from $CDAG_i$ by the Garbage collection method. The nodes without children are removed from the $CDAG_i$ (except for leaves). If during garbage removal, any new removal condition is seen, it is entered into the LRC. The removal conditions are applied to the SM and the $CDAG_i$.

**ALGORITHM 17:** The method to garbaging collection from $CDAG_i$

**GarbageCollectionCDAG**(**CDAG** $cdag$)
$flag \leftarrow false$
**for** $i \leftarrow cdag.lenght - 1$ $downto$ $1$ **do**
   **foreach** $node\ n$ **in** $column\ i\ of\ the\ cdag$
     **if**($n.left = null$ **and** $n.right = null$) **then**
       $Remove\ n.cluse\ from\ the\ SM\ and\ remove\ n\ from\ the\ cdag$
       $flag \leftarrow true$
     **end**
**end** // end of for
**Garbage collection**($SM, LRC$)
**Return** $flag$
**end**

$T(n) = $ length of CDAG $= \binom{n}{3} = \dfrac{n^3 - 3n^2 + 2n}{6} = O(n^3)$

**ALGORITHM 18:** The method to generating CDAG

// root $\in \{CM[1][column], 1 \leq column \leq 8\}$, root $= (l_1 \vee l_2 \vee l_3)$,
// $l_1 \in \{x_1, \bar{x}_1\}, l_2 \in \{x_2, \bar{x}_2\}, l_3 \in \{x_3, \bar{x}_3\}$
**GeneratingCDAG**(**Clause** $root$)
   $CDAG \leftarrow null$
   $LRC[2n][2n] \leftarrow PRC$
  **GeneratingSM**($column$)
  **if**(**MatrixIsValid**($SM$) $= true$) **then**
    $CDAG.root \leftarrow root$
    **for** $i \leftarrow 4$ $to$ $n$ **do**
      $CDAG_1 \leftarrow CDAG$
      $CDAG_2 \leftarrow CDAG$
      $c_1 \leftarrow (l_1 \vee l_2 \vee x_i)$
      $c_2 \leftarrow (l_1 \vee l_2 \vee \bar{x}_i)$;
      **if**($c_1 \in SM$) **then**
        **InsertToCDAGc**($c_1, CDAG_1$)
      **if**($c_2 \in SM$) **then**
        **InsertToCDAGc̄** ($c_2, CDAG_2$)
      $CDAG \leftarrow$ **Merge**($CDAG_1, CDAG_2$)
      **if**($CDAG \neq null$) **then**
        $state \leftarrow$ **GarbagecollectionCDAG**($CDAG$)
        **if**($state = true$) **then**
          **if**(**MatrixIsValid**($SM$) $= true$) **then**
            **GeneratingCDAG**($root$)
          **else if**(**MatrixIsValid**($SM$) $= true$) **then**
            **Return** $false$
        **end** // end of if(state)
      **end** //end of if($CDAG \neq null$)
      **else if** *(CDAG=null)* **then**
        **Return** $false$
    **end** // end of for(i)
  **end** // end of if (MatrixIsValid($SM$) $= true$)
  **else if** (**MatrixIsValid**($SM$) $= false$) **then**
    **Return** $false$
**Return** true
**end** //end of algorithm

$T(n) = \sum_{4}^{n}(n^6 + n^6) \times$ The count of clauses of the CM $= n \times n^6 \times 8 \times \binom{n}{3} = n \times n^6 \times 8 \times \dfrac{n^3 - 3n^2 + 2n}{6} = n \times n^6 \times n^3 = O(n^{10})$

If a node is removed from the $CDAG_i$ by the garbage removal, the construction of the CDAG restarts from stage 1 with the resulting SM from the Garbage collection method.

**Note:**
According to the algorithm in construction of the final CDAG, no node is removed from the CDAG as according to the algorithm in case of any removal, , the construction stages would restart from the scratch.

LEMMA 16. Each CDAG created by the offered algorithm is equivalent to a compatible binary tree.

PROOF. The proof is completed using the mathematic induction method. In this induction, the method for building a compatible binary tree equivalent to the CDAG, is explained.

**The induction base:**

If section $l_4$ contains part $x_4$, this part is supposed as the schema tree $t_4$. The clauses of the schema tree $t_4$ are compatible with each other according to variable $x_4$. If section $l_4$ contains part $\bar{x}_4$, this part is taken as schema tree $\bar{t}_4$. The clauses of the schema tree $\bar{t}_4$ are compatible with each other according to variable $\bar{x}_4$. Due to the structure of the CDAG, section $l_4$ incorporates at least one of the parts $x_4$ or $\bar{x}_4$ . The schema tree $T_4$ is generated as follows (see Fig. 8):

$$T_4 = t_4 \cup \bar{t}_4$$

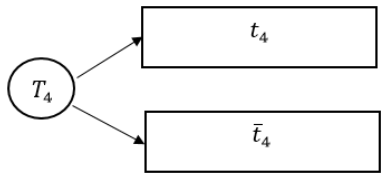

**Fig. 8. Schema tree $T_4$**

The schema tree $T_4$ consists of two completely separated sets. Worth noting that no clause is deleted from section $l_4$ and the clauses of section $l_4$ locate either in $t_4$ or $\bar{t}_4$.

If section $l_5$ contains section $x_5$, then a copy from schema tree $T_4$ is placed in front of section $x_5$ , By processing the copy of schema tree $T_4$ the schema tree $t_5$ is yielded according to the following method.

**The processing method of $T_4$ :**

$T_4$ is scanned from the last column to the first one, In this scan, if there are clauses relating to $l_5$ in the current column, the clauses containing variable $x_5$ are reserved and their parent or parents will be kept as the nodes of next stages (Which exist due to the REMOVAL LEMMA 2). Due to the REMOVAL LEMMA 2, in the current column, clauses containing variable $\bar{x}_5$ may also exist which are removed from the copy of schema tree $T_4$. If in the clauses of current column, there is no $l_5$ literal, without processing those clauses, their parents will be kept as the nodes of next processing stages. By having the task of scanning finished the schema tree $t_5$ is achieved. The clauses of schema tree $t_5$ are compatible relative to variable $x_5$.

If the section $l_5$ contains section $\bar{x}_5$ , a copy from schema tree $T_4$ is placed in front of the section $\bar{x}_5$ , By processing the copy of schema tree $T_4$, schema tree $\bar{t}_5$ is yielded according to the following method.

**The processing method of $T_4$ :**

$T_4$ is scanned from the last column to the first one, In this scan, if there are clauses relating to $l_5$ in the current column, the clauses containing variable $\bar{x}_5$ are reserved and their parent or parents will be kept as the nodes of next stages (Which exist due to the REMOVAL LEMMA 2). Due to the REMOVAL LEMMA 2, in the current column, clauses containing variable $x_5$ may also exist which are removed from the copy of schema tree $T_4$. If in the clauses of current column, there is no $l_5$ literal, without processing those clauses, their parents will be kept as the nodes of next processing stages. After the scanning task, finishing , schema tree $\bar{t}_5$ is achieved. The clauses of schema tree $\bar{t}_5$ are compatible relative to variable $\bar{x}_5$.

Due to the structure of the CDAG, section $l_5$ incorporates at least one of the parts $x_5$ or $\bar{x}_5$ . Schema tree $T_5$ is generated as follows ( see Fig. 9):

$$T_5 = t_5 \cup \bar{t}_5$$

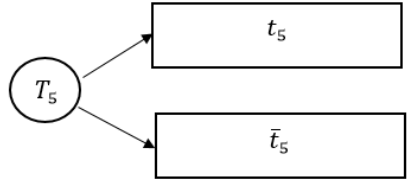

**Fig. 9. Schema tree $T_5$**

The schema tree $T_5$ consists of two completely separated sets. Worth noting that no clause is deleted from section $l_5$ and the clauses of section $l_5$ locate either in $t_5$ or $\bar{t}_5$.

The clauses of the routes in schema tree $T_5$ are compatible relative to variables $x_5, \bar{x}_5, x_4, \bar{x}_4$.

**The induction hypothesis:**

Schema tree $T_k$ exists in which the clauses over the routes of this schema tree are compatible relative to variables $x_k, \bar{x}_k, \ldots, x_5, \bar{x}_5, x_4, \bar{x}_4$.

## The induction mandate:

In (k+1)$_{th}$ stage, schema tree T$_{k+1}$ is created.

If section $l_{k+1}$ contains section $x_{k+1}$, a copy from the schema tree $T_k$ is placed in front of section $x_{k+1}$, By processing the copy of schema tree $T_k$, the schema tree $t_{k+1}$ is yielded according to the following method.

## The processing method of $T_k$:

$T_k$ is scanned from the last column to the first one, In this scan, if there exist clauses relating to $l_{k+1}$ in the current column, the clauses containing variable $x_{k+1}$ are reserved and their parent or parents will be kept as the nodes of next stages (Which exist due to the REMOVAL LEMMA 2). Due to the REMOVAL LEMMA 2, in the current column, clauses containing variable $\bar{x}_{k+1}$ may also exist which are removed from the copy of schema tree $T_k$. If in the clauses of current column, there is no $l_{k+1}$ literal, , their parents will be kept as the nodes of next processing stages without processing those clauses. By having the task of scanning finished, schema tree $t_{k+1}$ is achieved. The clauses of schema tree $t_{k+1}$ are compatible relative to variable $x_{k+1}$.

If section $l_{k+1}$ contains section $\bar{x}_{k+1}$, a copy from schema tree $T_k$ is placed in front of section $\bar{x}_{k+1}$, By processing the copy of schema tree $T_k$, schema tree $\bar{t}_{k+1}$ is yielded following the below method.

## The processing method of $T_k$:

$T_k$ is scanned from the last column to the first one, . In this scan, if there are clauses relating to $l_{k+1}$ in the current column, the clauses containing variable $\bar{x}_{k+1}$ are reserved and their parent or parents will be kept as the nodes of next stages (Which exist due to the REMOVAL LEMMA 2). Due to the REMOVAL LEMMA 2, in the current column, clauses containing variable $x_{k+1}$ may also exist which are removed from the copy of schema tree $T_k$. If in the clauses of current column, there is no $l_{k+1}$ literal, without processing those clauses, their parents will be kept as the nodes of next processing stages. By having the task of scanning finished, the schema tree $\bar{t}_{k+1}$ is achieved. The clauses of schema tree $\bar{t}_{k+1}$ are compatible relative to variable $\bar{x}_{k+1}$.

Due to structure of the CDAG, section $l_{k+1}$ incorporates at least one of the parts $x_{k+1}$ or $\bar{x}_{k+1}$. Schema tree $T_{k+1}$ is generated as follows (see Fig. 10):

$$T_{k+1} = t_{k+1} \cup \bar{t}_{k+1}$$

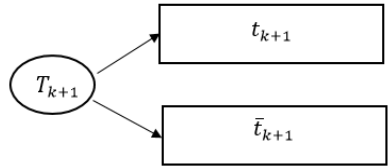

**Fig. 10. Schema tree $T_{k+1}$**

The schema tree $T_{k+1}$ consists of two completely separated sets. Worth noting that no clause is deleted from section $l_{k+1}$ and the clauses of section $l_{k+1}$ locate either in $t_{k+1}$ or $\bar{t}_{k+1}$.

The clauses of the routes of schema tree $T_{k+1}$ are compatible relative to variables $x_{k+1}, \bar{x}_{k+1}, \ldots, x_5, \bar{x}_5, x_4, \bar{x}_4$.

If the above-mentioned procedure continue until stage n, a binary tree with depth $\binom{n}{3}$ would be generated. Apparently, the length of the path from the root node to the leaf nodes in this tree would be also $\binom{n}{3}$. All clauses in a path would be consistent with each other because they have been correctly separated from each other. According to LEMMA 5, each path is actually a Clause_Set and according to LEMMA 13, the result value of $\varphi$ based on the Clause_Set would be true.

## 4 Algorithm

The proposed algorithm attempts to generate at least one of the CDAGs of the problem. If the algorithm can achieve this goal, the result value of the problem $\varphi$ would be true, otherwise, it would be false.

**ALGORITHM 18:** The main algorithm

```
//Input is a 3 − CNF − SAT like φ = c_1 ∧ ... ∧ c_m
//Output is "true" or "false"
sort all clauses of φ
GenerateCM()
Subtracting φ from CM( )
PRC[2n][2n] ← 0;
Garbage collection(CM, PRC)
if(MatrixIsValid(CM) = true) then
  for i ← 1 to 8 do
    if(CM[1][i] ≠ null) then
      state ← GeneratingCDAG(CM[1][i])
      if (state = true) then
        Output("true")
        Exit()
      end
      else if (state = false) then
          CM[1][i] ← null;
          FindRemovalConditions(CM, PRC)
          Garbage collection(CM)
          if(MatrixIsValid(CM) = false) then
            Output("false")
            Exit( )
          end
      end  // end of else
    end //  end of If(CM[1][i] ≠ null)
  end //end of for
  Output("false")
  Exit()
end //end of if(MatrixIsValid(CM) = true)
else if(MatrixIsValid(CM) = false)  then
    Output("false")
    Exit( )
end
end //end of algorithm
```

$T(n) = 8 \times T(n)$ of Generating CDAG $= 8 \times n^{10} = O(n^{10})$

LEMMA 17. If the problem is satisfiable, the algorithm returns true, and consequently if the algorithm returns true, the problem $\varphi$ is satisfiable.

PROOF. a) If the result value of the problem $\varphi$ is true, there would be at least one assignment in truth table, that makes the result value of the problem $\varphi$ as true. Also there would be a string $w$ in strings' table, which also makes the problem $\varphi$ as true. Therefore, none of the clauses in clause_set(w) includes the removal condition and will remain in the CM and the SM. According to the fact that all clauses in Clause_Set(w) are consistent with each other, the algorithm will create a CDAG based on $Clause\_Set(w)$ It is enough to choose the corresponding clause to the first index in the Clause_Set(w) e.g., $(x_1 \lor x_2 \lor x_3)$, as the root. In the following stages such as stage $i$ $(4 \leq i \leq n)$, because all clauses are resulted from combination of $(x_1 \lor x_2 \lor x_3)$ and clauses of the CDAG, are consistent with each other, and they exist in the CM and the SM, the resulted clauses (from combination) will be inserted into the CDAG, and therefore the CDAG will be generated.

b) The algorithm will return true as its result, if a CDAG is generated. In this case, according to LEMMA 16, the result value of the problem $\varphi$ would be true.

## 5 Sample

$$\varphi = (\bar{x}_1 \vee \bar{x}_2 \vee \bar{x}_3) \wedge (\bar{x}_2 \vee \bar{x}_3 \vee x_4) \wedge (\bar{x}_2 \vee \bar{x}_3 \vee \bar{x}_4) \wedge (x_1 \vee \bar{x}_2 \vee x_5) \wedge (\bar{x}_2 \vee x_3 \vee \bar{x}_5) \wedge (\bar{x}_1 \vee \bar{x}_2 \vee \bar{x}_6)$$

### 5.1 GenerateCM then ⬚⬚⬚⬚⬚⬚⬚⬚⬚⬚⬚ $\varphi$ ⬚⬚⬚⬚ CM (see Fig. 11)

| $x_1 \vee x_2 \vee x_3$ | $x_1 \vee x_2 \vee \bar{x}_3$ | $x_1 \vee \bar{x}_2 \vee x_3$ | $x_1 \vee \bar{x}_2 \vee \bar{x}_3$ | $\bar{x}_1 \vee x_2 \vee x_3$ | $\bar{x}_1 \vee x_2 \vee \bar{x}_3$ | $\bar{x}_1 \vee \bar{x}_2 \vee x_3$ | $\bar{x}_1 \vee \bar{x}_2 \vee \bar{x}_3$ |
|---|---|---|---|---|---|---|---|
| $x_1 \vee x_2 \vee x_4$ | $x_1 \vee x_2 \vee \bar{x}_4$ | $x_1 \vee \bar{x}_2 \vee x_4$ | $x_1 \vee \bar{x}_2 \vee \bar{x}_4$ | $\bar{x}_1 \vee x_2 \vee x_4$ | $\bar{x}_1 \vee x_2 \vee \bar{x}_4$ | $\bar{x}_1 \vee \bar{x}_2 \vee x_4$ | $\bar{x}_1 \vee \bar{x}_2 \vee \bar{x}_4$ |
| $x_1 \vee x_2 \vee x_5$ | $x_1 \vee x_2 \vee \bar{x}_5$ | $x_1 \vee \bar{x}_2 \vee x_5$ | $x_1 \vee \bar{x}_2 \vee \bar{x}_5$ | $\bar{x}_1 \vee x_2 \vee x_5$ | $\bar{x}_1 \vee x_2 \vee \bar{x}_5$ | $\bar{x}_1 \vee \bar{x}_2 \vee x_5$ | $\bar{x}_1 \vee \bar{x}_2 \vee \bar{x}_5$ |
| $x_1 \vee x_2 \vee x_6$ | $x_1 \vee x_2 \vee \bar{x}_6$ | $x_1 \vee \bar{x}_2 \vee x_6$ | $x_1 \vee \bar{x}_2 \vee \bar{x}_6$ | $\bar{x}_1 \vee x_2 \vee x_6$ | $\bar{x}_1 \vee x_2 \vee \bar{x}_6$ | $\bar{x}_1 \vee \bar{x}_2 \vee x_6$ | $\bar{x}_1 \vee \bar{x}_2 \vee \bar{x}_6$ |
| $x_1 \vee x_3 \vee x_4$ | $x_1 \vee x_3 \vee \bar{x}_4$ | $x_1 \vee \bar{x}_3 \vee x_4$ | $x_1 \vee \bar{x}_3 \vee \bar{x}_4$ | $\bar{x}_1 \vee x_3 \vee x_4$ | $\bar{x}_1 \vee x_3 \vee \bar{x}_4$ | $\bar{x}_1 \vee \bar{x}_3 \vee x_4$ | $\bar{x}_1 \vee \bar{x}_3 \vee \bar{x}_4$ |
| $x_1 \vee x_3 \vee x_5$ | $x_1 \vee x_3 \vee \bar{x}_5$ | $x_1 \vee \bar{x}_3 \vee x_5$ | $x_1 \vee \bar{x}_3 \vee \bar{x}_5$ | $\bar{x}_1 \vee x_3 \vee x_5$ | $\bar{x}_1 \vee x_3 \vee \bar{x}_5$ | $\bar{x}_1 \vee \bar{x}_3 \vee x_5$ | $\bar{x}_1 \vee \bar{x}_3 \vee \bar{x}_5$ |
| $x_1 \vee x_3 \vee x_6$ | $x_1 \vee x_3 \vee \bar{x}_6$ | $x_1 \vee \bar{x}_3 \vee x_6$ | $x_1 \vee \bar{x}_3 \vee \bar{x}_6$ | $\bar{x}_1 \vee x_3 \vee x_6$ | $\bar{x}_1 \vee x_3 \vee \bar{x}_6$ | $\bar{x}_1 \vee \bar{x}_3 \vee x_6$ | $\bar{x}_1 \vee \bar{x}_3 \vee \bar{x}_6$ |
| $x_1 \vee x_4 \vee x_5$ | $x_1 \vee x_4 \vee \bar{x}_5$ | $x_1 \vee \bar{x}_4 \vee x_5$ | $x_1 \vee \bar{x}_4 \vee \bar{x}_5$ | $\bar{x}_1 \vee x_4 \vee x_5$ | $\bar{x}_1 \vee x_4 \vee \bar{x}_5$ | $\bar{x}_1 \vee \bar{x}_4 \vee x_5$ | $\bar{x}_1 \vee \bar{x}_4 \vee \bar{x}_5$ |
| $x_1 \vee x_4 \vee x_6$ | $x_1 \vee x_4 \vee \bar{x}_6$ | $x_1 \vee \bar{x}_4 \vee x_6$ | $x_1 \vee \bar{x}_4 \vee \bar{x}_6$ | $\bar{x}_1 \vee x_4 \vee x_6$ | $\bar{x}_1 \vee x_4 \vee \bar{x}_6$ | $\bar{x}_1 \vee \bar{x}_4 \vee x_6$ | $\bar{x}_1 \vee \bar{x}_4 \vee \bar{x}_6$ |
| $x_1 \vee x_5 \vee x_6$ | $x_1 \vee x_5 \vee \bar{x}_6$ | $x_1 \vee \bar{x}_5 \vee x_6$ | $x_1 \vee \bar{x}_5 \vee \bar{x}_6$ | $\bar{x}_1 \vee x_5 \vee x_6$ | $\bar{x}_1 \vee x_5 \vee \bar{x}_6$ | $\bar{x}_1 \vee \bar{x}_5 \vee x_6$ | $\bar{x}_1 \vee \bar{x}_5 \vee \bar{x}_6$ |
| $x_2 \vee x_3 \vee x_4$ | $x_2 \vee x_3 \vee \bar{x}_4$ | $x_2 \vee \bar{x}_3 \vee x_4$ | $x_2 \vee \bar{x}_3 \vee \bar{x}_4$ | $\bar{x}_2 \vee x_3 \vee x_4$ | $\bar{x}_2 \vee x_3 \vee \bar{x}_4$ | $\bar{x}_2 \vee \bar{x}_3 \vee x_4$ | $\bar{x}_2 \vee \bar{x}_3 \vee \bar{x}_4$ |
| $x_2 \vee x_3 \vee x_5$ | $x_2 \vee x_3 \vee \bar{x}_5$ | $x_2 \vee \bar{x}_3 \vee x_5$ | $x_2 \vee \bar{x}_3 \vee \bar{x}_5$ | $\bar{x}_2 \vee x_3 \vee x_5$ | $\bar{x}_2 \vee x_3 \vee \bar{x}_5$ | $\bar{x}_2 \vee \bar{x}_3 \vee x_5$ | $\bar{x}_2 \vee \bar{x}_3 \vee \bar{x}_5$ |
| $x_2 \vee x_3 \vee x_6$ | $x_2 \vee x_3 \vee \bar{x}_6$ | $x_2 \vee \bar{x}_3 \vee x_6$ | $x_2 \vee \bar{x}_3 \vee \bar{x}_6$ | $\bar{x}_2 \vee x_3 \vee x_6$ | $\bar{x}_2 \vee x_3 \vee \bar{x}_6$ | $\bar{x}_2 \vee \bar{x}_3 \vee x_6$ | $\bar{x}_2 \vee \bar{x}_3 \vee \bar{x}_6$ |
| $x_2 \vee x_4 \vee x_5$ | $x_2 \vee x_4 \vee \bar{x}_5$ | $x_2 \vee \bar{x}_4 \vee x_5$ | $x_2 \vee \bar{x}_4 \vee \bar{x}_5$ | $\bar{x}_2 \vee x_4 \vee x_5$ | $\bar{x}_2 \vee x_4 \vee \bar{x}_5$ | $\bar{x}_2 \vee \bar{x}_4 \vee x_5$ | $\bar{x}_2 \vee \bar{x}_4 \vee \bar{x}_5$ |
| $x_2 \vee x_4 \vee x_6$ | $x_2 \vee x_4 \vee \bar{x}_6$ | $x_2 \vee \bar{x}_4 \vee x_6$ | $x_2 \vee \bar{x}_4 \vee \bar{x}_6$ | $\bar{x}_2 \vee x_4 \vee x_6$ | $\bar{x}_2 \vee x_4 \vee \bar{x}_6$ | $\bar{x}_2 \vee \bar{x}_4 \vee x_6$ | $\bar{x}_2 \vee \bar{x}_4 \vee \bar{x}_6$ |
| $x_2 \vee x_5 \vee x_6$ | $x_2 \vee x_5 \vee \bar{x}_6$ | $x_2 \vee \bar{x}_5 \vee x_6$ | $x_2 \vee \bar{x}_5 \vee \bar{x}_6$ | $\bar{x}_2 \vee x_5 \vee x_6$ | $\bar{x}_2 \vee x_5 \vee \bar{x}_6$ | $\bar{x}_2 \vee \bar{x}_5 \vee x_6$ | $\bar{x}_2 \vee \bar{x}_5 \vee \bar{x}_6$ |
| $x_3 \vee x_4 \vee x_5$ | $x_3 \vee x_4 \vee \bar{x}_5$ | $x_3 \vee \bar{x}_4 \vee x_5$ | $x_3 \vee \bar{x}_4 \vee \bar{x}_5$ | $\bar{x}_3 \vee x_4 \vee x_5$ | $\bar{x}_3 \vee x_4 \vee \bar{x}_5$ | $\bar{x}_3 \vee \bar{x}_4 \vee x_5$ | $\bar{x}_3 \vee \bar{x}_4 \vee \bar{x}_5$ |
| $x_3 \vee x_4 \vee x_6$ | $x_3 \vee x_4 \vee \bar{x}_6$ | $x_3 \vee \bar{x}_4 \vee x_6$ | $x_3 \vee \bar{x}_4 \vee \bar{x}_6$ | $\bar{x}_3 \vee x_4 \vee x_6$ | $\bar{x}_3 \vee x_4 \vee \bar{x}_6$ | $\bar{x}_3 \vee \bar{x}_4 \vee x_6$ | $\bar{x}_3 \vee \bar{x}_4 \vee \bar{x}_6$ |
| $x_3 \vee x_5 \vee x_6$ | $x_3 \vee x_5 \vee \bar{x}_6$ | $x_3 \vee \bar{x}_5 \vee x_6$ | $x_3 \vee \bar{x}_5 \vee \bar{x}_6$ | $\bar{x}_3 \vee x_5 \vee x_6$ | $\bar{x}_3 \vee x_5 \vee \bar{x}_6$ | $\bar{x}_3 \vee \bar{x}_5 \vee x_6$ | $\bar{x}_3 \vee \bar{x}_5 \vee \bar{x}_6$ |
| $x_4 \vee x_5 \vee x_6$ | $x_4 \vee x_5 \vee \bar{x}_6$ | $x_4 \vee \bar{x}_5 \vee x_6$ | $x_4 \vee \bar{x}_5 \vee \bar{x}_6$ | $\bar{x}_4 \vee x_5 \vee x_6$ | $\bar{x}_4 \vee x_5 \vee \bar{x}_6$ | $\bar{x}_4 \vee \bar{x}_5 \vee x_6$ | $\bar{x}_4 \vee \bar{x}_5 \vee \bar{x}_6$ |

| |
|---|
| Removal of clauses due to the subtracting $\varphi$ from CM |
| Removal of clauses due to the FindRemovalConditions($CM, PRC$), removal condition is $\bar{x}_2 \vee \bar{x}_3$ |
| Removal of clauses due to the FindRemovalConditions($CM, PRC$), removal condition is $\bar{x}_2 \vee \bar{x}_5$ |
| Removal of clauses due to the FindRemovalConditions($CM, PRC$), removal condition is $x_1 \vee \bar{x}_2$ |

Fig. 11. CM of $\varphi$

### 5.2 GeneratingSM(1) then Garbage Collection(SM) (see Fig. 12)

| $x_1 \vee x_2 \vee x_3$ | $x_1 \vee x_2 \vee \bar{x}_3$ | $x_1 \vee \bar{x}_2 \vee x_3$ | $x_1 \vee \bar{x}_2 \vee \bar{x}_3$ | $\bar{x}_1 \vee x_2 \vee x_3$ | $\bar{x}_1 \vee x_2 \vee \bar{x}_3$ | $\bar{x}_1 \vee \bar{x}_2 \vee x_3$ | $\bar{x}_1 \vee \bar{x}_2 \vee \bar{x}_3$ |
|---|---|---|---|---|---|---|---|
| $x_1 \vee x_2 \vee x_4$ | $x_1 \vee x_2 \vee \bar{x}_4$ | $x_1 \vee \bar{x}_2 \vee x_4$ | $x_1 \vee \bar{x}_2 \vee \bar{x}_4$ | $\bar{x}_1 \vee x_2 \vee x_4$ | $\bar{x}_1 \vee x_2 \vee \bar{x}_4$ | $\bar{x}_1 \vee \bar{x}_2 \vee x_4$ | $\bar{x}_1 \vee \bar{x}_2 \vee \bar{x}_4$ |
| $x_1 \vee x_2 \vee x_5$ | $x_1 \vee x_2 \vee \bar{x}_5$ | $x_1 \vee \bar{x}_2 \vee x_5$ | $x_1 \vee \bar{x}_2 \vee \bar{x}_5$ | $\bar{x}_1 \vee x_2 \vee x_5$ | $\bar{x}_1 \vee x_2 \vee \bar{x}_5$ | $\bar{x}_1 \vee \bar{x}_2 \vee x_5$ | $\bar{x}_1 \vee \bar{x}_2 \vee \bar{x}_5$ |
| $x_1 \vee x_2 \vee x_6$ | $x_1 \vee x_2 \vee \bar{x}_6$ | $x_1 \vee \bar{x}_2 \vee x_6$ | $x_1 \vee \bar{x}_2 \vee \bar{x}_6$ | $\bar{x}_1 \vee x_2 \vee x_6$ | $\bar{x}_1 \vee x_2 \vee \bar{x}_6$ | $\bar{x}_1 \vee \bar{x}_2 \vee x_6$ | $\bar{x}_1 \vee \bar{x}_2 \vee \bar{x}_6$ |
| $x_1 \vee x_3 \vee x_4$ | $x_1 \vee x_3 \vee \bar{x}_4$ | $x_1 \vee \bar{x}_3 \vee x_4$ | $x_1 \vee \bar{x}_3 \vee \bar{x}_4$ | $\bar{x}_1 \vee x_3 \vee x_4$ | $\bar{x}_1 \vee x_3 \vee \bar{x}_4$ | $\bar{x}_1 \vee \bar{x}_3 \vee x_4$ | $\bar{x}_1 \vee \bar{x}_3 \vee \bar{x}_4$ |
| $x_1 \vee x_3 \vee x_5$ | $x_1 \vee x_3 \vee \bar{x}_5$ | $x_1 \vee \bar{x}_3 \vee x_5$ | $x_1 \vee \bar{x}_3 \vee \bar{x}_5$ | $\bar{x}_1 \vee x_3 \vee x_5$ | $\bar{x}_1 \vee x_3 \vee \bar{x}_5$ | $\bar{x}_1 \vee \bar{x}_3 \vee x_5$ | $\bar{x}_1 \vee \bar{x}_3 \vee \bar{x}_5$ |
| $x_1 \vee x_3 \vee x_6$ | $x_1 \vee x_3 \vee \bar{x}_6$ | $x_1 \vee \bar{x}_3 \vee x_6$ | $x_1 \vee \bar{x}_3 \vee \bar{x}_6$ | $\bar{x}_1 \vee x_3 \vee x_6$ | $\bar{x}_1 \vee x_3 \vee \bar{x}_6$ | $\bar{x}_1 \vee \bar{x}_3 \vee x_6$ | $\bar{x}_1 \vee \bar{x}_3 \vee \bar{x}_6$ |
| $x_1 \vee x_4 \vee x_5$ | $x_1 \vee x_4 \vee \bar{x}_5$ | $x_1 \vee \bar{x}_4 \vee x_5$ | $x_1 \vee \bar{x}_4 \vee \bar{x}_5$ | $\bar{x}_1 \vee x_4 \vee x_5$ | $\bar{x}_1 \vee x_4 \vee \bar{x}_5$ | $\bar{x}_1 \vee \bar{x}_4 \vee x_5$ | $\bar{x}_1 \vee \bar{x}_4 \vee \bar{x}_5$ |
| $x_1 \vee x_4 \vee x_6$ | $x_1 \vee x_4 \vee \bar{x}_6$ | $x_1 \vee \bar{x}_4 \vee x_6$ | $x_1 \vee \bar{x}_4 \vee \bar{x}_6$ | $\bar{x}_1 \vee x_4 \vee x_6$ | $\bar{x}_1 \vee x_4 \vee \bar{x}_6$ | $\bar{x}_1 \vee \bar{x}_4 \vee x_6$ | $\bar{x}_1 \vee \bar{x}_4 \vee \bar{x}_6$ |
| $x_1 \vee x_5 \vee x_6$ | $x_1 \vee x_5 \vee \bar{x}_6$ | $x_1 \vee \bar{x}_5 \vee x_6$ | $x_1 \vee \bar{x}_5 \vee \bar{x}_6$ | $\bar{x}_1 \vee x_5 \vee x_6$ | $\bar{x}_1 \vee x_5 \vee \bar{x}_6$ | $\bar{x}_1 \vee \bar{x}_5 \vee x_6$ | $\bar{x}_1 \vee \bar{x}_5 \vee \bar{x}_6$ |
| $x_2 \vee x_3 \vee x_4$ | $x_2 \vee x_3 \vee \bar{x}_4$ | $x_2 \vee \bar{x}_3 \vee x_4$ | $x_2 \vee \bar{x}_3 \vee \bar{x}_4$ | $\bar{x}_2 \vee x_3 \vee x_4$ | $\bar{x}_2 \vee x_3 \vee \bar{x}_4$ | $\bar{x}_2 \vee \bar{x}_3 \vee x_4$ | $\bar{x}_2 \vee \bar{x}_3 \vee \bar{x}_4$ |
| $x_2 \vee x_3 \vee x_5$ | $x_2 \vee x_3 \vee \bar{x}_5$ | $x_2 \vee \bar{x}_3 \vee x_5$ | $x_2 \vee \bar{x}_3 \vee \bar{x}_5$ | $\bar{x}_2 \vee x_3 \vee x_5$ | $\bar{x}_2 \vee x_3 \vee \bar{x}_5$ | $\bar{x}_2 \vee \bar{x}_3 \vee x_5$ | $\bar{x}_2 \vee \bar{x}_3 \vee \bar{x}_5$ |
| $x_2 \vee x_3 \vee x_6$ | $x_2 \vee x_3 \vee \bar{x}_6$ | $x_2 \vee \bar{x}_3 \vee x_6$ | $x_2 \vee \bar{x}_3 \vee \bar{x}_6$ | $\bar{x}_2 \vee x_3 \vee x_6$ | $\bar{x}_2 \vee x_3 \vee \bar{x}_6$ | $\bar{x}_2 \vee \bar{x}_3 \vee x_6$ | $\bar{x}_2 \vee \bar{x}_3 \vee \bar{x}_6$ |
| $x_2 \vee x_4 \vee x_5$ | $x_2 \vee x_4 \vee \bar{x}_5$ | $x_2 \vee \bar{x}_4 \vee x_5$ | $x_2 \vee \bar{x}_4 \vee \bar{x}_5$ | $\bar{x}_2 \vee x_4 \vee x_5$ | $\bar{x}_2 \vee x_4 \vee \bar{x}_5$ | $\bar{x}_2 \vee \bar{x}_4 \vee x_5$ | $\bar{x}_2 \vee \bar{x}_4 \vee \bar{x}_5$ |
| $x_2 \vee x_4 \vee x_6$ | $x_2 \vee x_4 \vee \bar{x}_6$ | $x_2 \vee \bar{x}_4 \vee x_6$ | $x_2 \vee \bar{x}_4 \vee \bar{x}_6$ | $\bar{x}_2 \vee x_4 \vee x_6$ | $\bar{x}_2 \vee x_4 \vee \bar{x}_6$ | $\bar{x}_2 \vee \bar{x}_4 \vee x_6$ | $\bar{x}_2 \vee \bar{x}_4 \vee \bar{x}_6$ |
| $x_2 \vee x_5 \vee x_6$ | $x_2 \vee x_5 \vee \bar{x}_6$ | $x_2 \vee \bar{x}_5 \vee x_6$ | $x_2 \vee \bar{x}_5 \vee \bar{x}_6$ | $\bar{x}_2 \vee x_5 \vee x_6$ | $\bar{x}_2 \vee x_5 \vee \bar{x}_6$ | $\bar{x}_2 \vee \bar{x}_5 \vee x_6$ | $\bar{x}_2 \vee \bar{x}_5 \vee \bar{x}_6$ |
| $x_3 \vee x_4 \vee x_5$ | $x_3 \vee x_4 \vee \bar{x}_5$ | $x_3 \vee \bar{x}_4 \vee x_5$ | $x_3 \vee \bar{x}_4 \vee \bar{x}_5$ | $\bar{x}_3 \vee x_4 \vee x_5$ | $\bar{x}_3 \vee x_4 \vee \bar{x}_5$ | $\bar{x}_3 \vee \bar{x}_4 \vee x_5$ | $\bar{x}_3 \vee \bar{x}_4 \vee \bar{x}_5$ |
| $x_3 \vee x_4 \vee x_6$ | $x_3 \vee x_4 \vee \bar{x}_6$ | $x_3 \vee \bar{x}_4 \vee x_6$ | $x_3 \vee \bar{x}_4 \vee \bar{x}_6$ | $\bar{x}_3 \vee x_4 \vee x_6$ | $\bar{x}_3 \vee x_4 \vee \bar{x}_6$ | $\bar{x}_3 \vee \bar{x}_4 \vee x_6$ | $\bar{x}_3 \vee \bar{x}_4 \vee \bar{x}_6$ |
| $x_3 \vee x_5 \vee x_6$ | $x_3 \vee x_5 \vee \bar{x}_6$ | $x_3 \vee \bar{x}_5 \vee x_6$ | $x_3 \vee \bar{x}_5 \vee \bar{x}_6$ | $\bar{x}_3 \vee x_5 \vee x_6$ | $\bar{x}_3 \vee x_5 \vee \bar{x}_6$ | $\bar{x}_3 \vee \bar{x}_5 \vee x_6$ | $\bar{x}_3 \vee \bar{x}_5 \vee \bar{x}_6$ |
| $x_4 \vee x_5 \vee x_6$ | $x_4 \vee x_5 \vee \bar{x}_6$ | $x_4 \vee \bar{x}_5 \vee x_6$ | $x_4 \vee \bar{x}_5 \vee \bar{x}_6$ | $\bar{x}_4 \vee x_5 \vee x_6$ | $\bar{x}_4 \vee x_5 \vee \bar{x}_6$ | $\bar{x}_4 \vee \bar{x}_5 \vee x_6$ | $\bar{x}_4 \vee \bar{x}_5 \vee \bar{x}_6$ |

| Legend |
|---|
| Removal of clauses due to the Generating $SM$ |
| Removal of clauses due to the FindRemovalConditions($SM, LRC$), removal condition is $\bar{x}_1$ |
| Removal of clauses due to the FindRemovalConditions($SM, LRC$), removal condition is $\bar{x}_2$ |
| Removal of clauses due to the FindRemovalConditions($SM, LRC$), removal condition is $\bar{x}_3$ |

**Fig. 12. SM of root $(x_1 \vee x_2 \vee x_3)$**

### 5.3 InsertToCDAGc$((x_1 \vee x_2 \vee x_4), CDAG_1)$ (see Fig. 13)

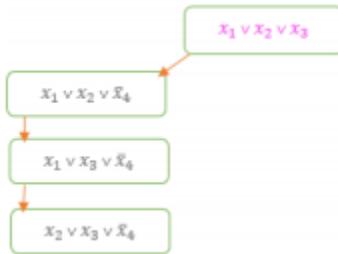

**Fig. 13. Result of InsertToCDAGc$((x_1 \vee x_2 \vee x_4), CDAG_1)$**

### 5.4 InsertToCDAG$\bar{c}$ $((x_1 \vee x_2 \vee \bar{x}_4), CDAG_2)$ (see Fig. 14)

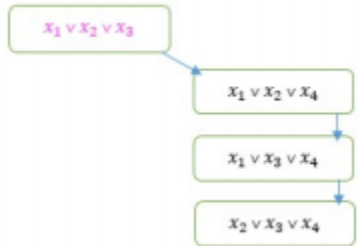

**Fig. 14. Result of InsertToCDAG$\bar{c}$$((x_1 \vee x_2 \vee \bar{x}_4), CDAG_2)$**

5.5 $CDAG \leftarrow$ **Merge**$(CDAG_1, CDAG_2)$(see Fig. 15)

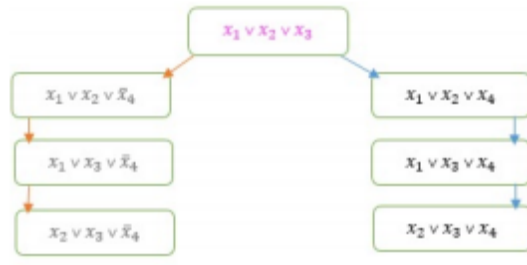

**Fig. 15. CDAG**

5.6 **InsertToCDAGc**$((x_1 \lor x_2 \lor x_5), CDAG_1)$ **(see Fig. 16)**

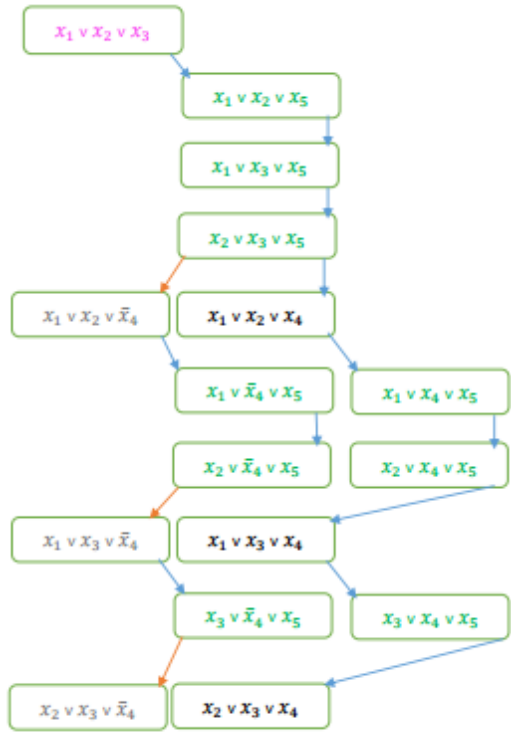

**Fig. 16. Result of InsertToCDAGc**$((x_1 \lor x_2 \lor x_5), CDAG_1)$

5.7 **InsertToCDAG$\bar{c}$** $((x_1 \lor x_2 \lor \bar{x}_5), CDAG_2)$ (see Fig. 17)

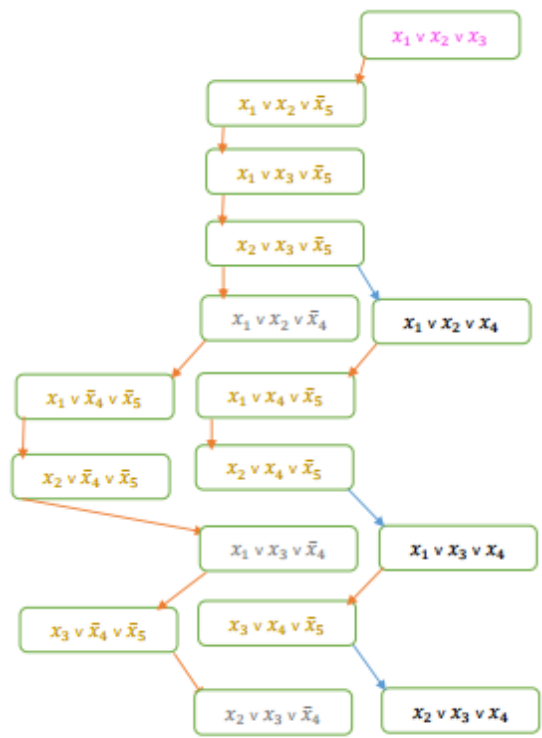

**Fig. 17. Result of InsertToCDAGc̄(($x_1 \lor x_2 \lor \bar{x}_5$), $CDAG_2$)**

**5.8** $CDAG \leftarrow$ **Merge**($CDAG_1, CDAG_2$) (see Fig. 18)

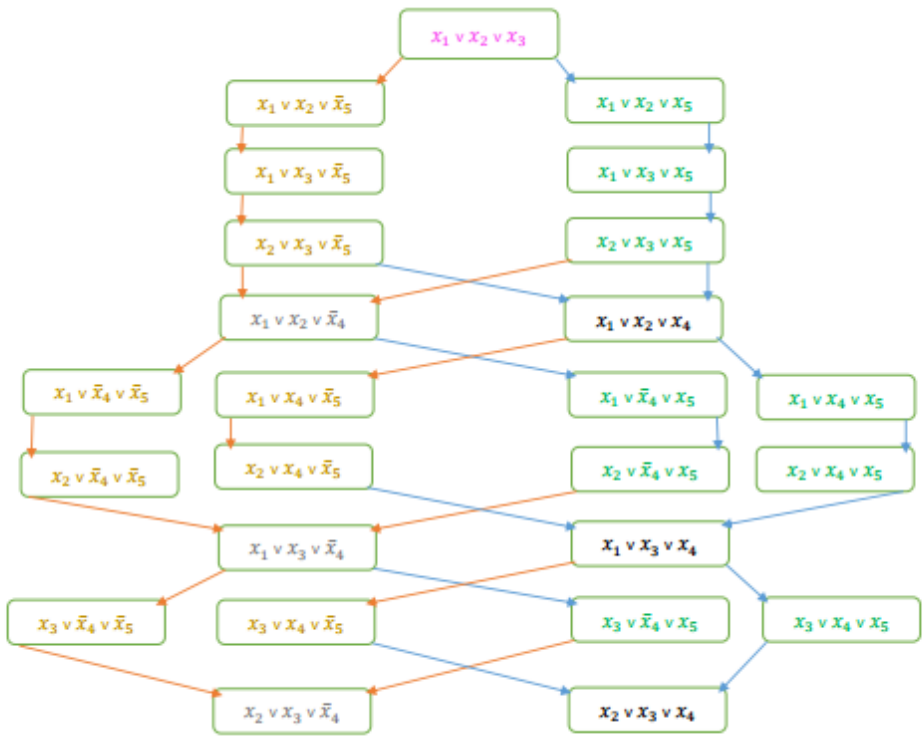

**Fig. 18. CDAG**

## 5.9 InsertToCDAGc$((x_1 \lor x_2 \lor x_6), CDAG_1)$ (see Fig. 19)

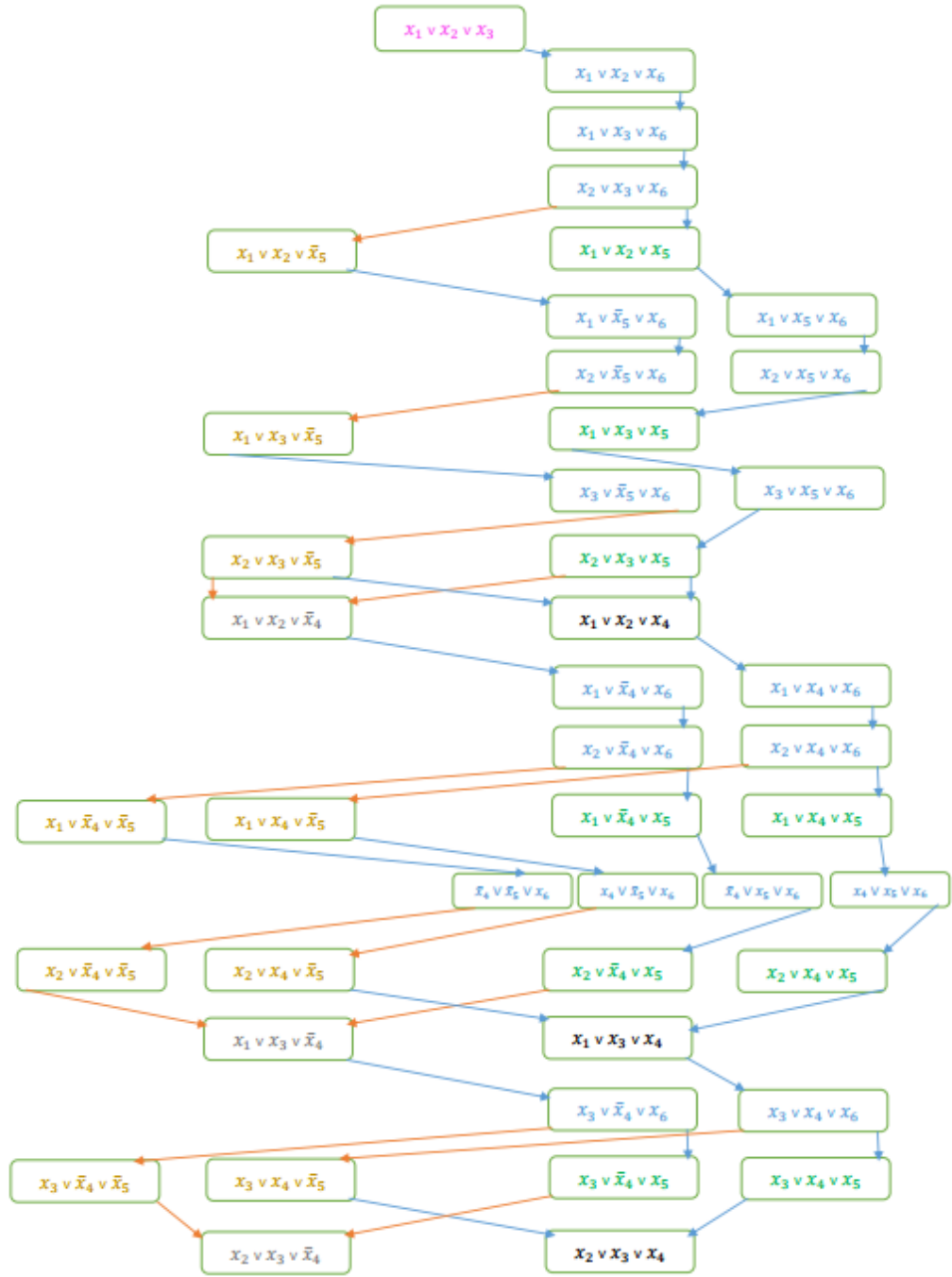

**Fig. 19. Result of InsertToCDAGc$((x_1 \lor x_2 \lor x_6), CDAG_1)$**

## 5.10 InsertToCDAGc̄ $((x_1 \lor x_2 \lor \bar{x}_6), CDAG_2)$ (see Fig. 20)

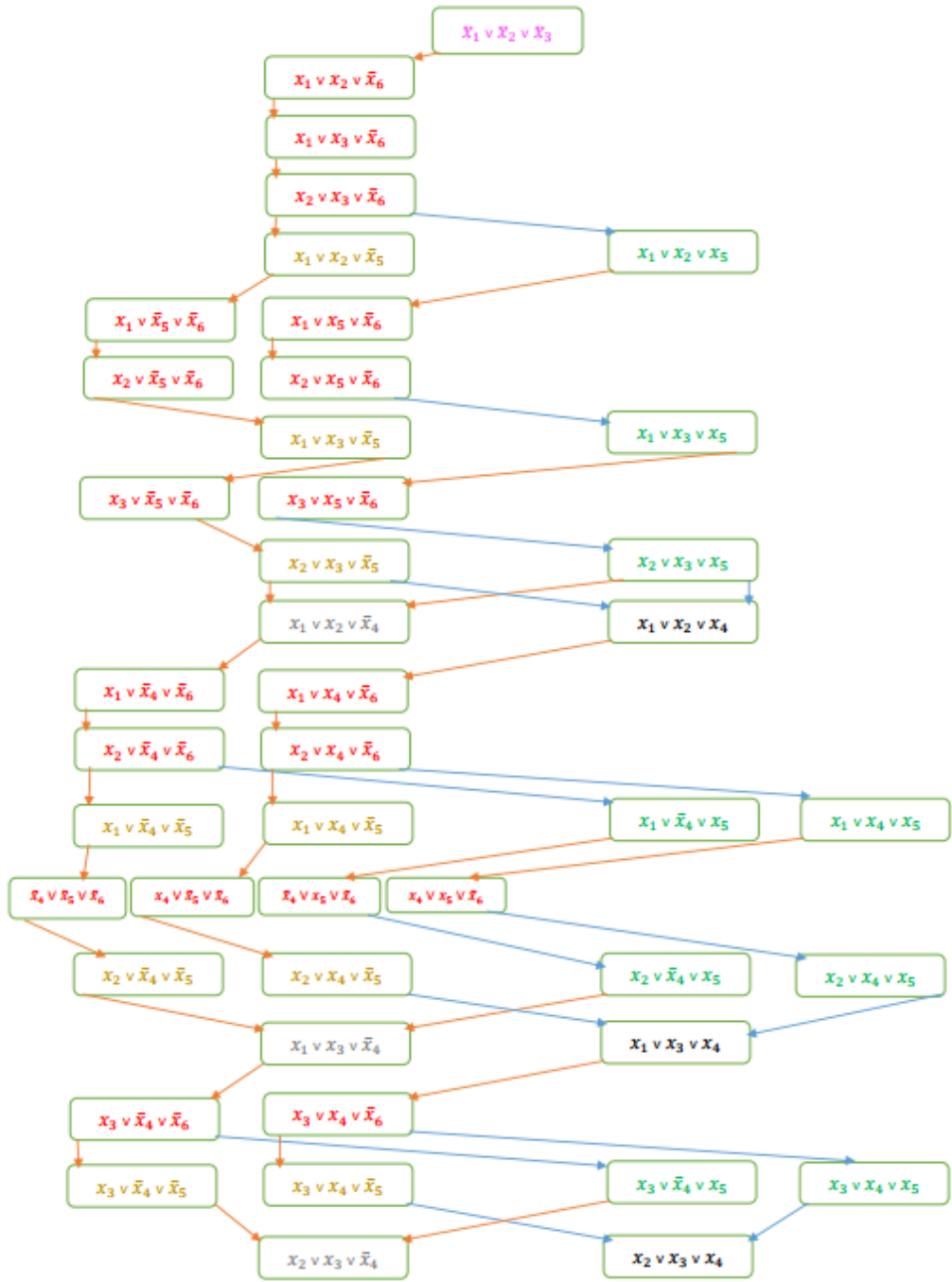

**Fig. 20.** Result of InsertToCDAG$\bar{c}$(($x_1 \vee x_2 \vee \bar{x}_6$), $CDAG_2$)

**5.11** $CDAG \leftarrow$ **Merge**($CDAG_1, CDAG_2$) (see Fig. 21)

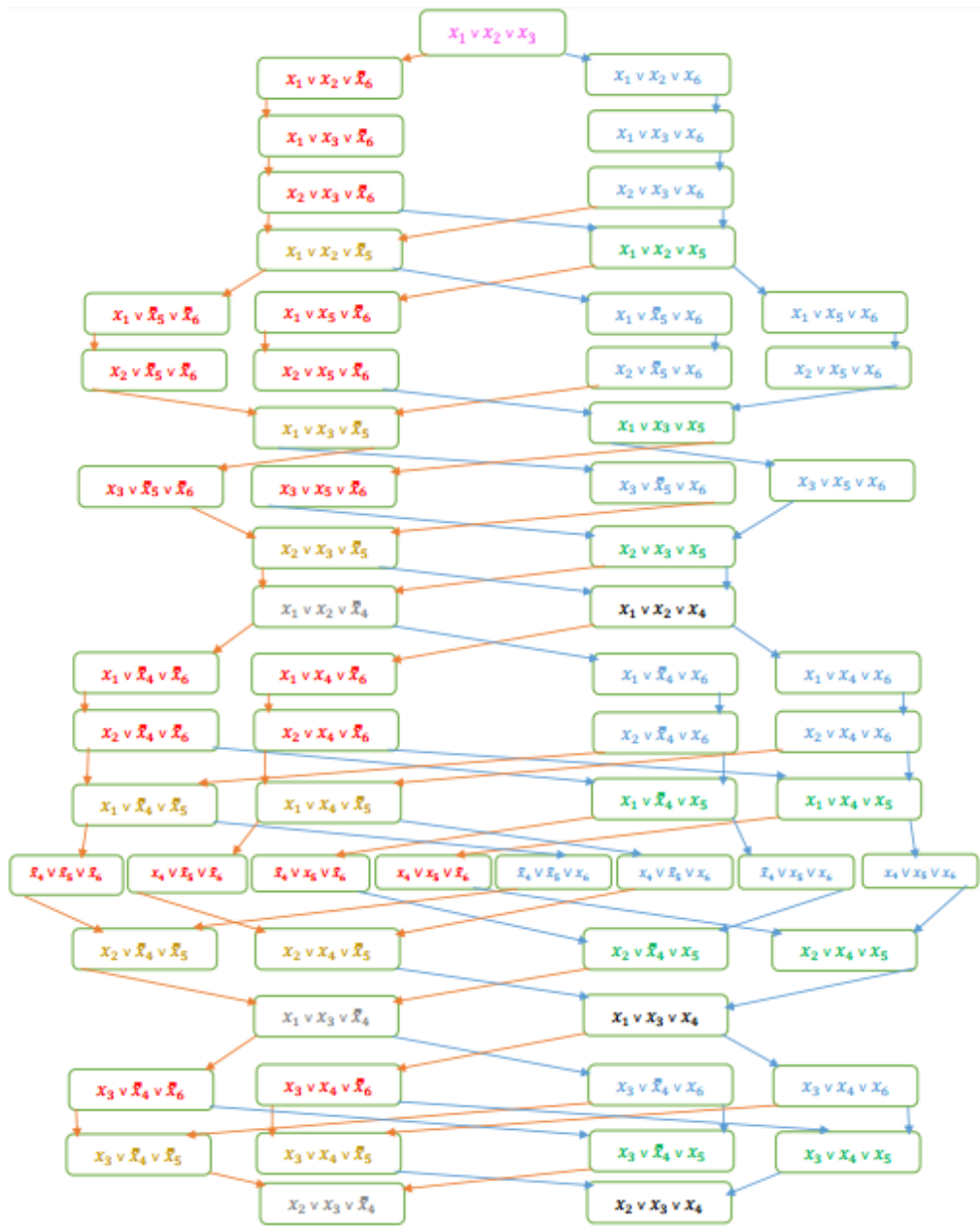

**Fig. 21. CDAG**

**5.12 For instance, a clause set is selected from the generated DAG in Figure 21 (see Fig. 22).**

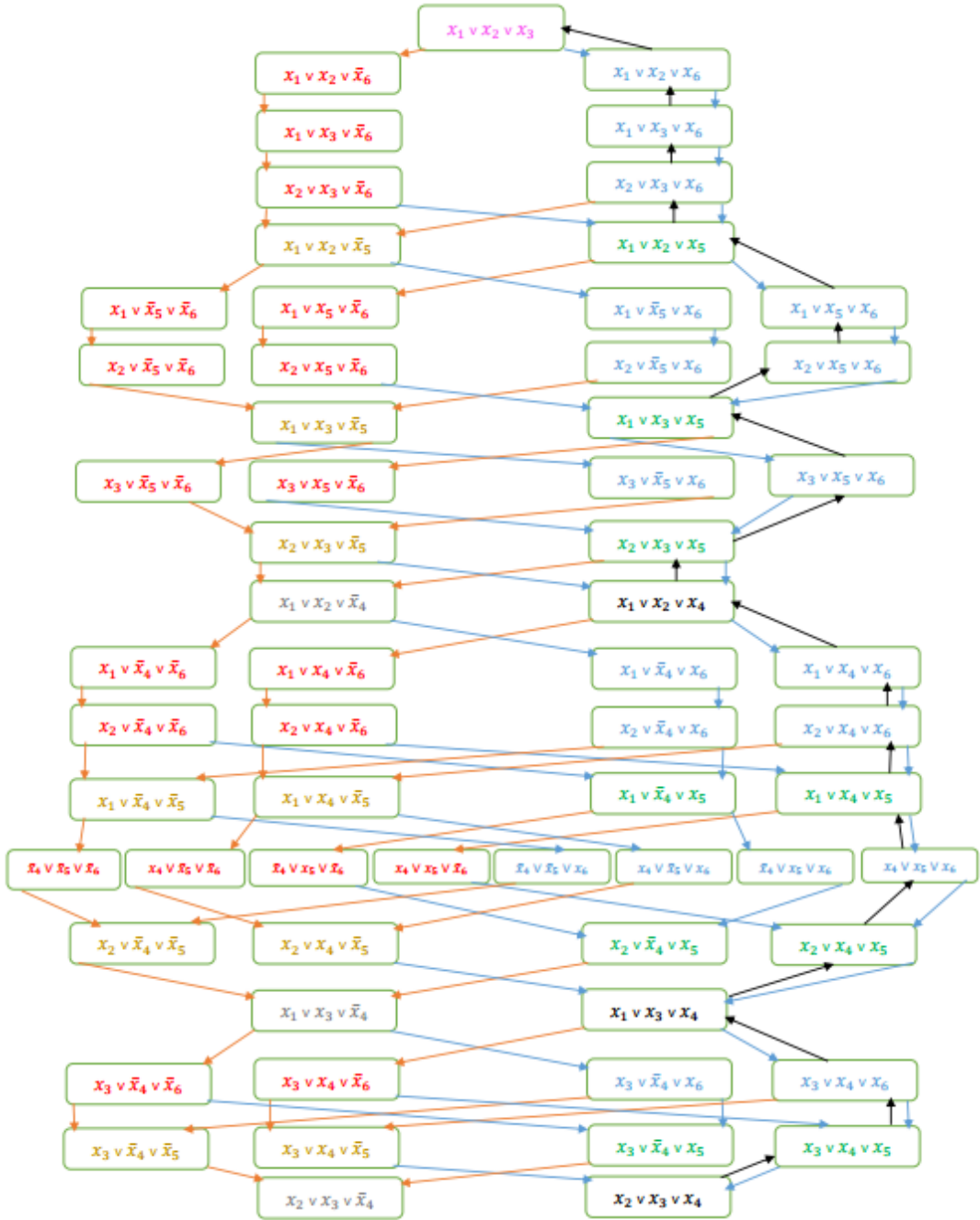

**Fig. 22. A Claus_Set**

**Clause_Set(w)** = {$(x_1 \vee x_2 \vee x_3), (x_1 \vee x_2 \vee x_4), (x_1 \vee x_2 \vee x_5), (x_1 \vee x_2 \vee x_6), (x_1 \vee x_3 \vee x_4), (x_1 \vee x_3 \vee x_5),$
$(x_1 \vee x_3 \vee x_6), (x_1 \vee x_4 \vee x_5), (x_1 \vee x_4 \vee x_6), (x_1 \vee x_5 \vee x_6), (x_2 \vee x_3 \vee x_4), (x_2 \vee x_3 \vee x_5), (x_2 \vee x_3 \vee x_6),$
$(x_2 \vee x_4 \vee x_5), (x_2 \vee x_4 \vee x_6), (x_2 \vee x_5 \vee x_6), (x_3 \vee x_4 \vee x_5), (x_3 \vee x_4 \vee x_6), (x_3 \vee x_5 \vee x_6), (x_4 \vee x_5 \vee x_6)$ }

$x_1 = false,$
$x_2 = false,$
$x_3 = false,$
$x_4 = false,$
$x_5 = false,$

$x_6 = false$

$\varphi = (\bar{x}_1 \vee \bar{x}_2 \vee \bar{x}_3) \wedge (\bar{x}_2 \vee \bar{x}_3 \vee x_4) \wedge (\bar{x}_2 \vee \bar{x}_3 \vee \bar{x}_4) \wedge (x_1 \vee \bar{x}_2 \vee x_5) \wedge (\bar{x}_2 \vee x_3 \vee \bar{x}_5) \wedge (\bar{x}_1 \vee \bar{x}_2 \vee \bar{x}_6)$

$Value((\bar{x}_1 \vee \bar{x}_2 \vee \bar{x}_3)) = (\overline{false} \vee \overline{false} \vee \overline{false}) = (true \vee true \vee true) = true$

$Value((\bar{x}_2 \vee \bar{x}_3 \vee x_4)) = (\overline{false} \vee \overline{false} \vee false) = (true \vee true \vee false) = true$

$Value((\bar{x}_2 \vee \bar{x}_3 \vee \bar{x}_4)) = (\overline{false} \vee \overline{false} \vee \overline{false}) = (true \vee true \vee true) = true$

$Value((x_1 \vee \bar{x}_2 \vee x_5)) = (false \vee \overline{false} \vee false) = (false \vee true \vee false) = true$

$Value((\bar{x}_2 \vee x_3 \vee \bar{x}_5)) = (\overline{false} \vee false \vee \overline{false}) = (true \vee false \vee true) = true$

$Value((\bar{x}_1 \vee \bar{x}_2 \vee \bar{x}_6)) = (\overline{false} \vee \overline{false} \vee \overline{false}) = (true \vee true \vee true) = true$

$Value(\varphi) = (true \wedge true \wedge true \wedge true \wedge true \wedge true) = true$

## 6 conclosion

The time-complexity of algorithm is $O(n^{10})$, so the problem 3-CNF-SAT is solved by engaging an algorithm with space complexity of $O(n^3)$ and time complexity of $O(n^{10})$. So that, result of P versus NP problem is P = NP.

## 7 Refrences